\documentclass[twocolumn,showpacs,nofootinbib,preprintnumbers,amsmath,amssymb,showkeys,superscriptaddress]{revtex4}

\usepackage[pdftex]{graphicx}  
\usepackage{color}
\usepackage{slashed}
\usepackage[normalem]{ulem}

\newcommand{\chiSB}{$\chi$SB}
\newcommand{\cb}{\overline c}

\newcommand \beq{\begin{eqnarray}}
\newcommand \eeq{\end{eqnarray}}

\begin{document}

\title{Small parameters in infrared quantum chromodynamics}

\author{Marcela Pel\'aez\vspace{.4cm}}%
\affiliation{%
Instituto de F\'{\i}sica, Facultad de Ingenier\'{\i}a, Universidad de
la Rep\'ublica, J. H. y Reissig 565, 11000 Montevideo, Uruguay.
\vspace{.1cm}}%
\author{Urko Reinosa}%
\affiliation{%
Centre de Physique Th\'eorique, Ecole Polytechnique, CNRS, Universit\'e Paris-Saclay, F-91128 Palaiseau, France.
\vspace{.1cm}}%
\author{Julien Serreau}%
\affiliation{APC, AstroParticule et Cosmologie, Universit\'e Paris Diderot, CNRS/IN2P3, CEA/Irfu, Observatoire de Paris, Sorbonne Paris Cit\'e, 10, rue Alice Domon et L\'eonie Duquet, 75205 Paris Cedex 13, France.\vspace{.1cm}}
\author{Matthieu Tissier}\affiliation{Laboratoire de Physique Th\'eorique de la Mati\`ere Condens\'ee, UPMC, CNRS UMR 7600, Sorbonne
Universit\'es, 4 Place Jussieu,75252 Paris Cedex 05, France.
\vspace{.1cm}}%
\author{Nicol\'as Wschebor}%
\affiliation{%
 Instituto de F\'{\i}sica, Facultad de Ingenier\'{\i}a, Universidad de
 la Rep\'ublica, J. H. y Reissig 565, 11000 Montevideo, Uruguay.
\vspace{.1cm}}%

\date{\today}

\begin{abstract}
  We study the long-distance properties of quantum chromodynamics in the Landau gauge in
  an expansion in powers of the three-gluon, four-gluon, and
  ghost-gluon couplings, but without expanding in the quark-gluon
  coupling. This is motivated by two observations. First, the
  gauge sector  is well-described by perturbation theory in the context of a phenomenological
  model with a massive gluon. Second, the quark-gluon coupling is significantly larger
  than those in the gauge sector at large
  distances. In order to resum the contributions of the remaining infinite set of QED-like diagrams, 
  we further expand the theory in $1/N_c$, where
  $N_c$ is the number of colors. At leading order, this double
  expansion leads to the well-known rainbow approximation for the quark propagator.  We 
  take advantage of the systematic expansion to get a
  renormalization-group improvement of the rainbow resummation. A
  simple numerical solution of the resulting coupled set of equations 
  reproduces
  the phenomenology of the spontaneous chiral symmetry breaking: for
  sufficiently large quark-gluon coupling constant, the constituent
  quark mass saturates when its valence mass approaches zero. We find
  very good agreement with lattice data for the scalar part of the
  propagator and explain why the vectorial part is poorly reproduced.
\end{abstract}

\maketitle

\section{Introduction}

The long-distance regime of quantum chromodynamics (QCD) is the arena
of several important phenomena. Of
utmost phenomenological relevance is the so-called spontaneous chiral
symmetry breaking (S\chiSB), which is responsible for the dramatic
increase of the running mass of the light quarks, from a few MeV to
roughly a third of the nucleon mass, when the renormalization-group
(RG) scale is lowered from a few GeV down to zero. This behavior is now
clearly established by lattice simulations, see e.g.
\cite{Bowman:2005vx,Oliveira:2016muq}, but its description within analytic
approaches remains a difficult problem. Indeed, this requires one to
control the theory in a regime where the couplings are large, or
even undefined, if one trusts standard perturbation theory. In fact,
it is widely believed that the whole infrared regime of QCD is
nonperturbative in nature and that its properties can be accessed only
through nonperturbative approaches, such as nonperturbative
renormalization group (NPRG), Schwinger-Dyson (SD) equations, the Hamiltonian formalism or
lattice simulations \cite{Schleifenbaum:2006bq,Quandt:2013wna,vonSmekal97,Alkofer00,Zwanziger01,Fischer03,Bloch03,Pawlowski:2003hq,Fischer:2004uk,Aguilar04,Boucaud06,Aguilar07,Aguilar08,Boucaud08,Fischer08,RodriguezQuintero10,Huber:2012kd,Sternbeck:2007ug,Cucchieri:2007rg,Cucchieri:2008fc,Sternbeck:2008mv,Cucchieri:2009zt,Bogolubsky:2009dc,Dudal:2010tf}.

On the analytical side, it is well understood that the physics of S\chiSB{} can be reproduced by
retaining a certain family of diagrams, the so-called rainbow
truncation (for classical references on the subject,  see
Refs.~\cite{Johnson:1964da,Maskawa:1974vs,Maskawa:1975hx,Miransky:1984ef,Atkinson:1988mv,Atkinson:1988mw,Bhagwat:2004hn};
some recent reviews are Refs.~\cite{Maris:2003vk,Roberts:2007jh}).  The corresponding truncation for two-body bound states, the so-called rainbow-ladder truncation, has been successfully applied to meson spectroscopy \cite{Maris:1999nt,Eichmann:2008ae}.  This appears naturally as the first nontrivial contribution in some nonperturbative approximation schemes, e.g., based on $n$-particle-irreducible techniques \cite{Sanchis-Alepuz:2015tha,Williams:2015cvx,Eichmann:2016yit}.  Note, however, that 
S\chiSB{}  requires a
sufficiently large coupling, as was first pointed out by Nambu and
Jona-Lasinio \cite{Nambu:1961tp,Nambu:1961fr}. Consequently, it
remains unclear why the particular family of rainbow diagrams should be retained while
some other diagrams are discarded.  Moreover, some modeling is usually necessary
for the gluon propagator and  the quark-gluon vertex.

A clue in order to explain the success of the rainbow truncation may be the following.
Recent works have shown that the dynamics in the gauge sector can be described by perturbative means within a massive
deformation of the standard Landau gauge QCD Lagrangian \cite{Tissier:2010ts,Tissier:2011ey,Pelaez:2013cpa,Reinosa:2017qtf}. This is motivated by thorough
studies of QCD correlation functions with lattice
simulations, the solutions of truncated SD and NPRG equations, as well as variational methods in the Hamiltonian formalism
\cite{Schleifenbaum:2006bq,Quandt:2013wna,vonSmekal97,Alkofer00,Zwanziger01,Fischer03,Bloch03,Pawlowski:2003hq,Fischer:2004uk,Aguilar04,Boucaud06,Aguilar07,Aguilar08,Boucaud08,Fischer08,RodriguezQuintero10,Huber:2012kd}. In the Landau gauge, the gluon propagator displays a saturation at small
momenta (the so-called decoupling --- or massive --- solution), while the ghost propagator presents a massless behaviour at vanishing
momentum
\cite{Sternbeck:2007ug,Cucchieri:2007rg,Cucchieri:2008fc,Sternbeck:2008mv,Cucchieri:2009zt,Bogolubsky:2009dc,Dudal:2010tf},
as in the bare theory. The physical origin of this massive behavior
for the gluons evades the usual perturbative treatment of the
theory. However, there are strong evidences which indicate that this
gluon mass is the major nonperturbative ingredient of the infrared
regime of Yang-Mills theory (for a recent general discussion on the topic, see \cite{Reinosa:2017qtf}). Indeed,
for what concerns pure Yang-Mills theories, it was shown in a series of articles
\cite{Tissier:2010ts,Tissier:2011ey,Pelaez:2013cpa} that one-loop calculations of two- and
three-points correlation functions in a simple extension of the Landau gauge Faddeev-Popov Lagrangian by means of a (phenomenologically motivated)
gluon mass compare quite well with lattice
simulations, with a maximal error ranging from 10\% to 20\% depending
on the correlation function. This is a particular case of the class of Curci-Ferrari (CF) Lagrangians \cite{Curci76}. This surprising result can be traced
back to the fact that, within this phenomenological model, the
interaction strength $\alpha_S$ remains moderate, even in the infrared
regime,\footnote{In particular, there exists RG
  schemes in which no Landau pole occurs \cite{Tissier:2011ey,Weber:2011nw}.} in agreement with lattice
simulations.

Similar studies were also performed with dynamical quarks
\cite{Pelaez:2014mxa,Pelaez:2015tba}: the gluon, ghost and quark
propagators, as well as the quark-gluon correlation function were
computed at one loop in the massive extension of Landau-gauge QCD. Most of the
correlation functions that could be compared with lattice simulations
showed the correct qualitative behaviors, with the noticeable exception
of the vectorial part of the quark propagator.\footnote{The reason for this mismatch can be traced back to the fact that,
in the Landau gauge and in the case of a massless gluon, the vectorial part of the quark self-energy at one loop vanishes identically,
see, for instance, \cite{Davydychev:2000rt}. In the presence of the gluon mass,
this contribution is abnormally small and comparable with the two-loop corrections.} However, for small
values of the bare quark masses, the quantitative comparison to lattice data
was less convincing in the quark sector.

\begin{figure}[tbp]
  \centering
  \includegraphics[width=\columnwidth]{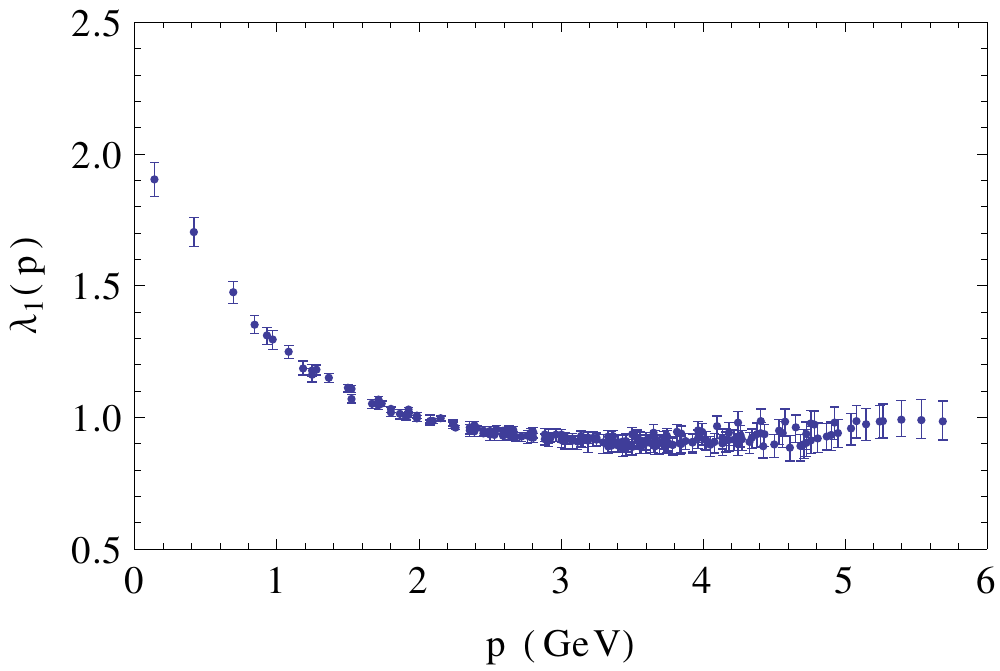}
  \caption{A measure of the relative of the quark-gluon and the ghost-gluon couplings. Figure generated from the lattice data of
  Ref.~\cite{Skullerud:2003qu}.\label{lambda1}}
\end{figure}

Again, lattice simulations give us an important clue for understanding
this poorer results in presence of dynamical quarks \cite{Oliveira:2016muq,Skullerud:2003qu}. Indeed, although equal in the ultraviolet,
the coupling constants of the different sectors of the theory differ significantly at long distances. Lattice simulations show that the coupling
in the quark sector is  two to three times larger in the infrared than the one in the gauge sector. This has also been observed in SD and NPRG
contexts \cite{Aguilar:2014lha,Braun:2014ata} as well as in the one-loop calculation of the quark-guon vertex of Ref.~\cite{Pelaez:2015tba}.
This is illustrated in Fig.~\ref{lambda1}, where
we show the ratio between the quark-gluon and ghost-gluon vertices in some kinematical configuration.  In this situation, a perturbative expansion in powers
of the quark-gluon coupling is questionable (recall that the relevant expansion parameter is proportional to the square of the coupling). Note that the fact that the quark-gluon coupling must be larger than the one observed in the gluonic sector is also in line with phenonomenological considerations \cite{Roberts:2007jh}: the coupling observed on the lattice in the gluonic
sector is too small to trigger the S\chiSB.

In this article, we propose to extend the work of \cite{Pelaez:2014mxa,Pelaez:2015tba} 
by taking into account the above observations. We treat the couplings in the gauge sector perturbatively while keeping all orders of the quark-gluon coupling. At leading
order, this reduces the set of diagrams to those appearing in an
Abelian theory. We further use an expansion in the number of colors $N_c$ \cite{'tHooft:1974hx} to obtain closed expressions for the associated correlation functions
(for a classical reference on
the validity of the large-$N_c$ limit in QCD, see, for instance, Ref.~\cite{Witten:1979kh}; for a recent numerical analysis of the question
see, for instance, Ref.~\cite{DeGrand:2016pur}). At leading order in $1/N_c$, this reduces to the
 rainbow-ladder diagrams.\footnote{ It is known that the large-$N_c$ limit coincides with the rainbow-ladder system of equations in the Nambu-Jona-Lasinio model \cite{Dmitrasinovic:1995cb,Nikolov:1996jj,Oertel:2000jp}. In QCD, because of the interactions in the gauge sector, the large-$N_c$ limit involves an infinite series of planar diagrams beyond those contributing to the rainbow-ladder approximation. An attempt to relate the large-$N_c$ limit and the rainbow-ladder approximation in QCD, using an effective gluon propagator, can be found in Ref.~\cite{Fischer:2007ze}.}

This is most welcome since this set of diagrams is known to capture the physics of S\chiSB{}
\cite{Johnson:1964da,Maskawa:1974vs,Maskawa:1975hx,Miransky:1984ef,Atkinson:1988mv,Atkinson:1988mw,Bhagwat:2004hn}.
The benefit of the present approach is that  this approximation is obtained in a controlled expansion that can be, in principle, systematically improved.
In particular, at leading order, the structure of the gluon propagator is determined by perturbation theory in the CF model.
Moreover, this allows for a consistent treatment of both the ultraviolet renormalization and the RG improvement of the rainbow-ladder approximation.

We solve
the resulting equations and show that they lead to a dramatic increase
of the running quark mass in the infrared and to a dynamically  
generated quark mass in the chiral limit.  At a
qualitative level, our results reproduce the expected feature that the
chiral symmetry breaking occurs for a sufficiently large
coupling. We show by an explicit comparison with lattice simulations
that our solution describes with precision the scalar component of the
quark propagator for various values of the bare masses (including
values close to the chiral limit). The vectorial component has
the right behavior in the ultraviolet regime but is not correctly
reproduced in the infrared, for reasons similar to the perturbative case
mentioned above. We stress that, even though one could, in principle, solve the complete set of equations that arise from our expansion scheme
at leading order, we use here, for simplicity, an ansatz for the running coupling. A complete treatment is deferred to a subsequent work.

The article is organized as follows. In
Sec.~\ref{sec_model}, we present the massive extension of Landau-gauge QCD and
we describe the double expansion in the couplings of the pure gauge sector and in $1/N_c$ in Sec.~\ref{sec_expansion}.
In Sec.~\ref{sec_calcul}, we write the equations which describe the resummation of the corresponding Feynman diagrams at leading order.
We implement the corresponding renormalization group improvement in Sec.~\ref{RG}. Finally, we solve the system of RG-improved integro-differential
equations for the quark propagator and compare our results with lattice data in Sec.~\ref{sec_compare}. We conclude in Sec.~\ref{sec_conclusion}.
Some technical material related to the RG improvement is gathered in an appendix.

\section{Massive Landau-gauge QCD}
\label{sec_model}

Let us start by giving a short review of the model. As is well-known
since the pioneering work of Gribov \cite{Gribov77} the Faddeev-Popov
procedure to fix the gauge in non-Abelian gauge theories is not
justified in the infrared regime, because of the so-called Gribov
ambiguity. To overcome this issue, Gribov \cite{Gribov77} and
Zwanziger \cite{Zwanziger89,Zwanziger92} have proposed to modify the
gauge-fixing procedure. Although this approach does not completely fix the Gribov ambiguity and requires taking into account many new auxiliary fields,
it has been applied with success to the determination of correlation functions (in its refined version \cite{Dudal08}) or to the study of the
deconfinement transition \cite{Dudal08,Canfora:2015yia}. Here instead, we follow the line initiated in Refs.~\cite{Tissier:2010ts} and use a more phenomenological approach which consists in adding a gluon mass term to the Faddeev-Popov action in the Landau gauge.\footnote{Such a massive extension has been discussed in relation with the Gribov problem in
Ref.~\cite{Serreau:2012cg}. We also mention that a related approach was developed in Ref.~\cite{Siringo:2015wtx}. This modifies the infrared behavior of the propagators in agreement with the findings of lattice simulations while maintaining the properties of standard perturbation theory in the
ultraviolet (including the renormalisability of the model). Also, this avoids the introduction of further auxiliary fields and leads to tractable analytical calculations \cite{Tissier:2011ey,Pelaez:2013cpa}. 
}
Following these considerations, we work with the QCD action, expressed
in Euclidean space, with the usual Landau gauge-fixing terms supplemented with a gluon mass term
\begin{equation}
  \label{eq_action}
  \begin{split}
      S=\int d^d&x\Bigg[\frac 14 F_{\mu\nu}^aF_{\mu\nu}^a+ih^a\partial_\mu
    A_\mu^a +\partial_\mu\cb^a(D_\mu c)^a\\
&+\frac 12 m_\Lambda^2 (A_\mu^a)^2
   + \sum_{i=1}^{N_f}\bar\psi_i(\slashed D + M_\Lambda)\psi_i  \Bigg].
  \end{split}
\end{equation}
The covariant derivatives applied to fields in the adjoint
($X$) and fundamental ($\psi$) representations read respectively
\begin{align*}
(D_{\mu}X)^a&=\partial_{\mu}X^a+g_\Lambda f^{abc}A_{\mu}^b X^c,\\
D_{\mu}\psi&=\partial_{\mu}\psi-ig_\Lambda A_{\mu}^a t^a \psi,
\end{align*}
with $f^{abc}$ the structure constants of the gauge group and $t^a$
the generators of the algebra in the fundamental representation. The
Euclidean Dirac matrices $\gamma$ satisfy
$\{\gamma_\mu,\gamma_\nu\}=2\delta_{\mu\nu}$,
$\slashed D=\gamma_\mu D_\mu$ and
$F_{\mu\nu}^a=\partial_\mu A_\nu^a -\partial_\nu A_\mu^a+ g_\Lambda
f^{abc}A_\mu^a A_\nu ^b$ is the field-strength tensor. Finally, the parameters $g_\Lambda$,
$M_\Lambda$ and $m_\Lambda$ are respectively the bare coupling
constant, quark mass and gluon mass, defined at some ultraviolet 
scale $\Lambda$. For simplicity, we only consider degenerate quark
masses, but the generalization to a more realistic case is
trivial. The previous action is standard, except for the gluon
mass. In actual perturbative calculations, this mass term appears
through a modified bare gluon propagator, which reads
\begin{equation}
  \label{eq_gluon_propag}
  G_{0,\mu\nu}^{ab}(p)=\delta^{ab}\frac
  1{p^2+m_\Lambda^2}\left(\delta_{\mu\nu}-\frac{p_\mu p_\nu}{p^2}\right).
\end{equation}

The gluon and ghost sectors of this model have been studied in
\cite{Tissier:2010ts,Tissier:2011ey,Pelaez:2013cpa} by using
perturbation theory. The quenched and unquenched two-point functions
for gluons and ghosts were calculated at one-loop order and compared
to the lattice simulations with an impressive agreement in view of the simplicity of the calculations. The
ghost-gluon and three-gluon vertices were also calculated and compared
rather well to lattice data.\footnote{Note however that the lattice
  data for three-point vertices have larger error bars than for
  propagators so that this test is less stringent. Very recently, more
  accurate lattice results for the three-gluon vertex have been announced
  \cite{Athenodorou:2016oyh,Boucaud:2017obn} but, for the moment,
  these results have not been compared to those of
  Ref.~\cite{Pelaez:2013cpa}.}  These perturbative calculations of correlations functions have been extended to finite temperature in
  Refs.~\cite{Reinosa:2013twa,Reinosa:2016iml}.  Also, physical observables, such as the
phase diagram and the behaviour of the Polyakov loop, were 
calculated with success \cite{Reinosa:2014ooa,Reinosa:2015oua}. In
some cases, two-loop calculations have been implemented and show an
improvement with respect to one-loop results
\cite{Reinosa:2014zta,Reinosa:2015gxn}. To summarize, there are strong evidences that
correlation functions in the gauge sector can be calculated
perturbatively with the model (\ref{eq_action}). The reason for that
is the absence of a Landau pole in the RG (for a
certain class of renormalization schemes) and the fact that the relevant
coupling in the ghost/gluon sector remains moderate even in the
infrared. In fact, it was shown in Ref.~\cite{Tissier:2011ey} that the
running expansion parameter is always smaller than $0.4$, and that this
rather large value is reached only in a small range of RG
scale.

The quark sector of QCD was also studied in
Refs.~\cite{Pelaez:2014mxa,Pelaez:2015tba} within the phenomenological model
(\ref{eq_action}) and we briefly discuss the main results obtained there. The
(renormalized) quark propagator $S$, can be parametrized as:
\begin{align}
  S(p)&=\left[-i A(p)\slashed p+B(p)\right]^{-1}=i \tilde A(p)\slashed p+\tilde B(p)\,,
\end{align}
where
\begin{align}
\tilde A(p) &=\frac{A(p)}{ A^2(p)p^2+B^2(p)}\,,\\
\tilde B(p) &=\frac{B(p)}{ A^2(p)p^2+B^2(p)}\,,
\end{align}
so that the tree-level propagator corresponds to $A=1$ and $B=M_\Lambda$. In
Ref.~\cite{Pelaez:2014mxa}, a one-loop calculation of the quark propagator
leads to a function $M(p)=B(p)/A(p)$ which compares
qualitatively well with lattice data when the bare quark mass is not too
small. In particular, there is an important enhancement of the running
quark mass in the infrared. However, when the bare quark mass
approaches the chiral limit, the mass function $M(p)$ goes to zero
and the spontaneous chiral symmetry breaking (S$\chi$SB) does not show
up. This is not surprising because since the works of Nambu and
Jona-Lasinio \cite{Nambu:1961tp,Nambu:1961fr}, S$\chi$SB is expected
to occur for couplings above a certain critical value. Such nonanalytic behavior cannot be captured at finite loop order.
A second disagreement of the results of Ref.~\cite{Pelaez:2014mxa} with
lattice data concerns the function $A(p)$, but its origin is much
less profound. As is well known, there is no one-loop
correction to the function $A(p)$ in the Landau gauge, when the
gluon mass is set to zero (see, for instance, Ref.~\cite{Davydychev:2000rt}). When the gluon mass is introduced, a
(finite) contribution to $A(p)$ is generated at one loop, which is, 
however, abnormally small and turns out to be of the same order as
two-loop corrections. In this situation, the one-loop approximation
is not justified and one would need to include two-loop
corrections. The latter have not been computed so far in the model
(\ref{eq_action}) but the plausibility of this scenario was tested in Ref.~\cite{Pelaez:2014mxa}, where the known results for the two-loop
contribution in the ultraviolet regime \cite{Gracey:2002yt} were
included to the analysis of the function $A$. This yielded a
good agreement with lattice data.  

Finally, the one-loop results for the quark-gluon vertex \cite{Pelaez:2015tba} are in qualitative agreement with
the lattice data for all scalar components and for all momentum configurations that have been simulated. Overall, the
agreement becomes poorer at very low momenta and is generally better for quantities that are not
sensitive to S$\chi$SB.

The main conclusion of such comparisons of one-loop perturbative results in the  phenomenological model (\ref{eq_action})
against lattice data is that the agreement is significantly better in the pure gauge sector than in the quark sector. 
This can be understood from the relative magnitudes of
the corresponding coupling constants. Of course, the running of the strong
coupling constant is universal at one and two loops in the ultraviolet
regime. However, this property is lost beyond two loops and also in a
mass-dependent scheme for momenta that are comparable to or smaller than the largest
mass in the problem.  For instance, as mentioned in the Introduction, a quantity that measures the
relative size of the quark-gluon coupling compared to the ghost-gluon
vertex is measured on the lattice \cite{Skullerud:2003qu} and is represented in
Fig.~\ref{lambda1}. One
observes that the quark-gluon coupling is significantly larger in the
infrared. Moreover, taking into account that the actual expansion parameter
of perturbation theory is proportional to the square of the
coupling, we conclude that the expansion parameter is about five times
larger in the quark sector than in the gluon/ghost sector.
The typical size of the latter being about a few tenths along the relevant momentum range \cite{Tissier:2011ey,Reinosa:2017qtf},
one concludes that the perturbative treatment of the quark-gluon vertices is not justified. In any case, the nontrivial phenomenon of S$\chi$SB{}
is beyond the reach of a purely perturbative analysis at any finite loop order.

\section{A new approximation scheme}
\label{sec_expansion}
 
\begin{figure}[tbp]
  \centering
  \includegraphics[width=\columnwidth]{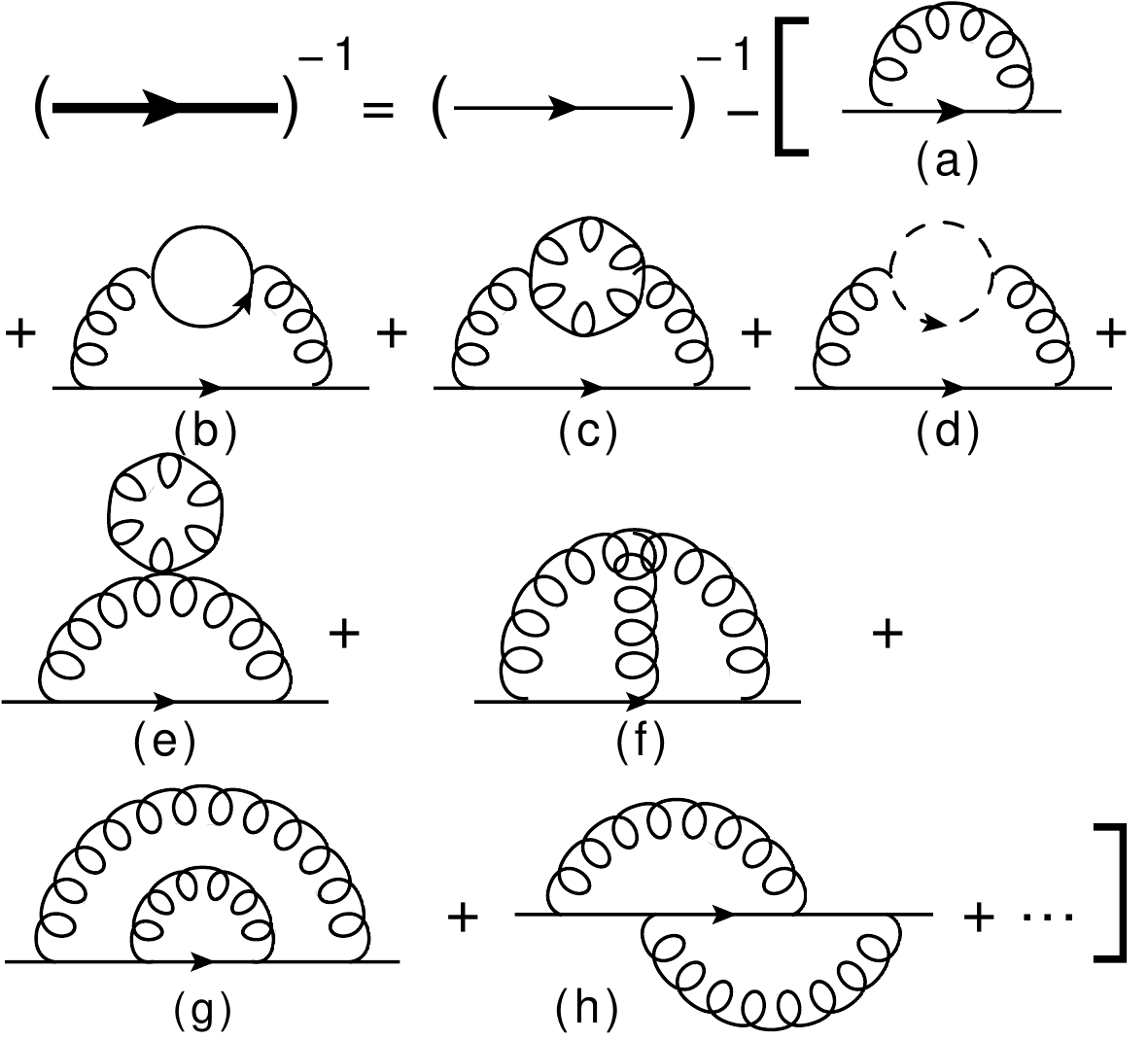}
  \caption{One and two loops Feynman diagrams contributing to the quark self-energy.\label{Fig:selfquarkTwoLoops}}
\end{figure}

 To overcome the problems of perturbation theory in the quark sector,
we propose an improved approximation scheme where the gluon/ghost
couplings (denoted by $g_g$) are treated perturbatively but
where all powers of the quark-gluon coupling (denoted by $g_q$)
are taken into account. We first discuss the example of the quark
self-energy, whose one- and two-loops diagrams are shown in
Fig~\ref{Fig:selfquarkTwoLoops}.
\begin{figure}[tbp]
  \centering
  \includegraphics[width=\columnwidth]{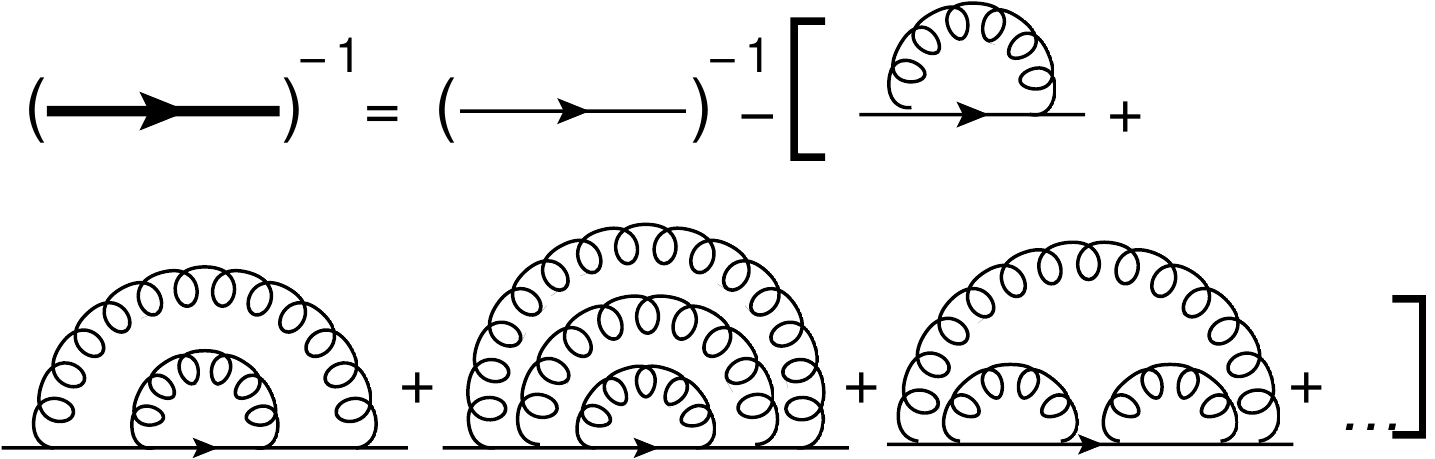}
  \caption{ Feynman
    diagrams with at most three loops contributing to the quark self-energy at leading order in a double expansion in large $N_c$ and
    small $g_g$. These are the rainbow diagrams of lowest order in
    quark-gluon coupling.\label{Fig:selfquarkRainbows}}
\end{figure}
Diagrams (c)--(f) can be ignored at leading order because
they are suppressed by one or two powers of $g_g$. More generally, neglecting diagrams with nonzero powers of $g_g$ leaves us with the infinite set of QED-like diagrams which, however, has no known closed analytic expression.
We further simplify the problem by organizing this set in powers of $1/N_c$ at fixed 't Hooft coupling $\lambda=g_q^2 N_c$, where $N_c$ is the number
of colors \cite{'tHooft:1974hx}. At leading order, only planar diagrams (i.e., with
quark lines on the border of the diagram) with no quark loop contribute. In
the example of Fig~\ref{Fig:selfquarkTwoLoops}, the diagrams (b) and
(h) are suppressed and the only diagrams left are (a) and (g). This
analysis can be generalized to all orders. The result is well-known:
only rainbow diagrams survive as represented in
Fig~\ref{Fig:selfquarkRainbows}.  This set of diagrams can be
resummed through an integral equation for the quark propagator which reads, diagrammatically, 
\begin{equation}
\includegraphics[width=.9\columnwidth]{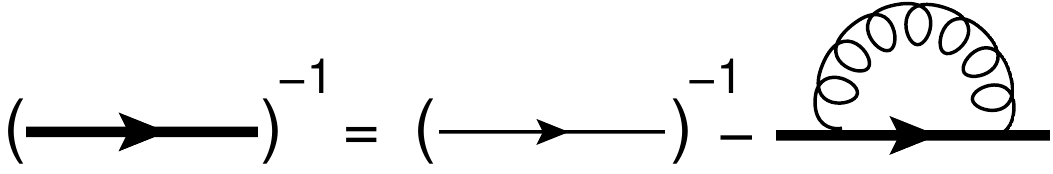},
  \label{Fig:eq_rainbow}
\end{equation}
where the thick line represents the (resummed) quark propagator at leading
order.  We can easily guess the predictions inferred
  from this set of diagrams in the ultraviolet. Indeed, the
  universality of the coupling constants and asymptotic freedom ensure that
  $g_g\sim g_q\ll 1$. In this limit, the quark self-energy is
  dominated by the contribution of the first diagram in the bracket of
  Fig.~\ref{Fig:selfquarkRainbows}. This observation is important
  because it ensures that the one-loop ultraviolet behavior is recovered in this
  approximation.\footnote{In practice, we shall keep the combinatorial factors of finite $N_c$ in order to preserve the one-loop exactness  of the approximation for any value of $N_c$.}

  The previous analysis can be generalized to any correlation
  function. To improve standard perturbation theory at $\ell$-loop
  order and take into account the fact that $g_q$ is significantly
  larger than $g_g$ in the infrared, write all diagrams of standard
  perturbation theory with up to $\ell$ loops, count the powers of
  $g_g$ and $1/N_c$ that appear in these diagrams and add all diagrams
  (with possibly more loops) with the same powers of $g_g$ and
  $1/N_c$. By construction, this set of diagrams reproduces the
  results of standard perturbation theory at $\ell$-loop order, but also
  reproduces, at leading order, the rainbow-ladder
  approximation. In what follows, we shall refer to this approximation
    scheme as the rainbow-improved (RI) loop expansion.

\begin{figure}[htbp]
  \centering
  \includegraphics[width=\columnwidth]{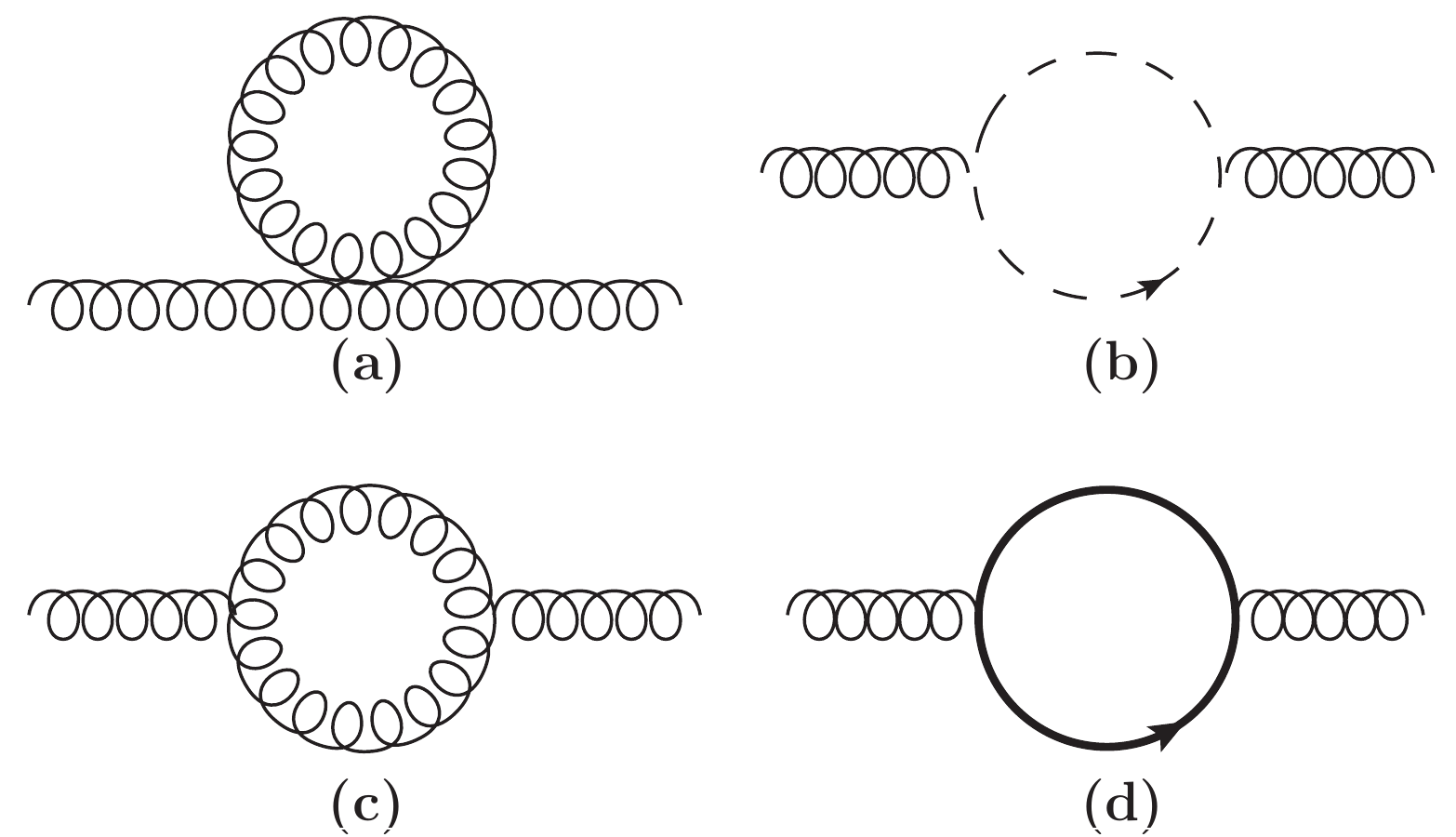}
  \includegraphics[width=4cm]{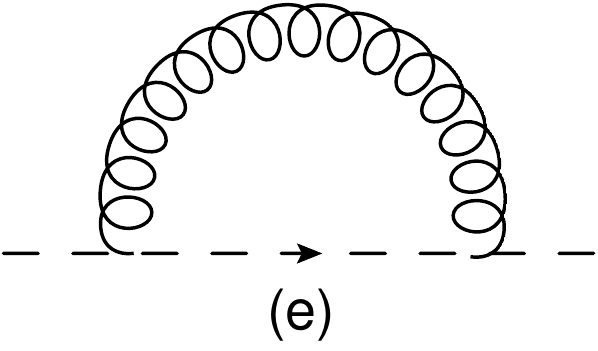}
  \caption{Diagrams contributing to the gluon (first two lines) and
    ghost (last line) self-energy at leading order in standard
    perturbation theory. \label{Fig:ghost-gluon}}
\end{figure}

  As a next example, we now discuss the cases of the gluon and ghost
  two-point self-energies at RI-one-loop order, depicted in Fig.~\ref{Fig:ghost-gluon}.  
 The standard one-loop structures in the pure gauge sector, i.e., diagrams
(a), (b), (c), and (e), are of order $g_g^2$, whereas the standard quark loop diagram is
of order $1/N_c$.
By inspection, we find that the set of diagrams with
the same powers of $g_g^2$ and $1/N_c$ are obtained by dressing the
quark propagator according to Fig.~\ref{Fig:selfquarkRainbows}, as represented by the thick line in diagram (d) of Figure~\ref{Fig:ghost-gluon}.

\begin{figure}[htbp]
  \centering
  \includegraphics[width=6cm]{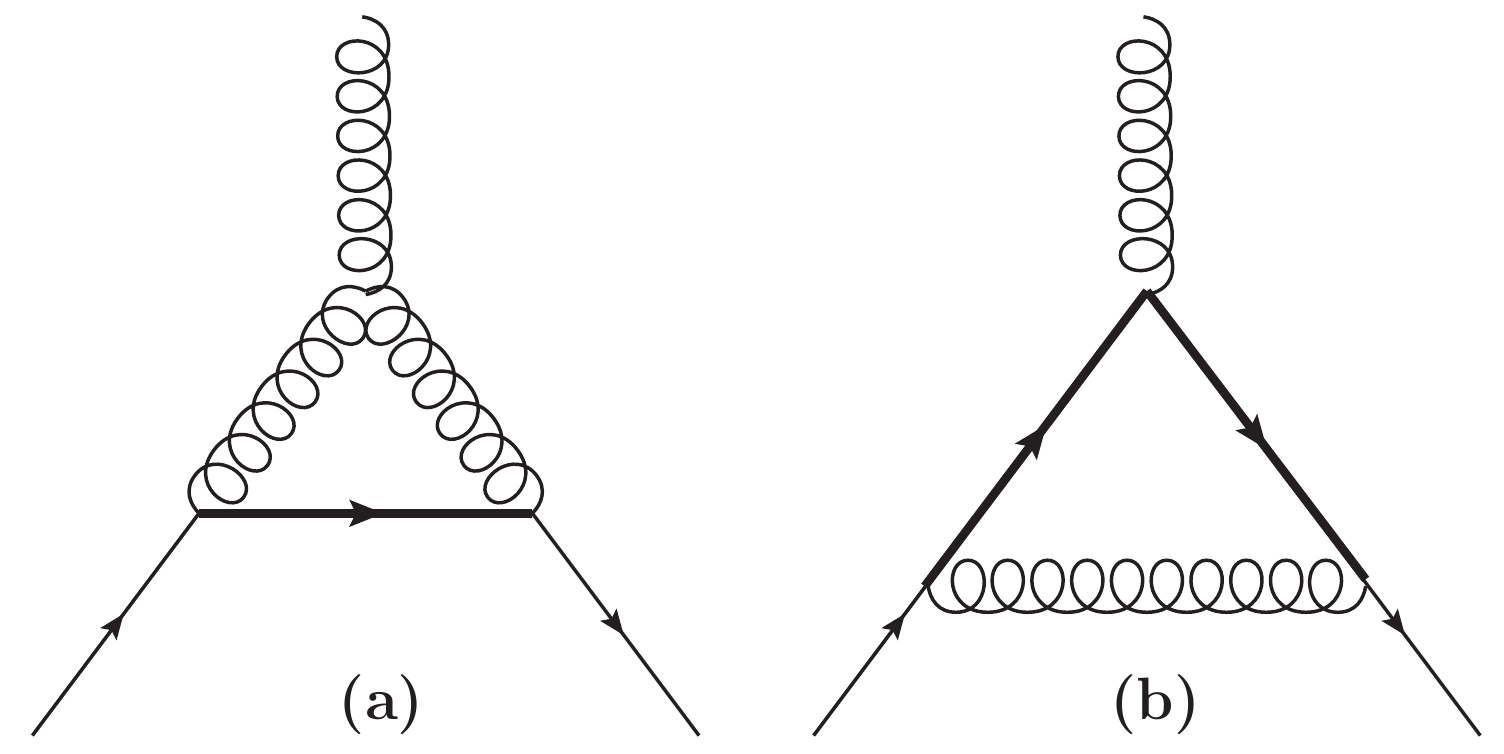}
  \caption{One-loop-order diagrams contributing to the quark-gluon 
    vertex. The standard one-loop structures are identical, with the resummed (thick) quark line replaced by the tree-level one. Diagram (a) involves a three-gluon coupling whereas diagram (b) is $1/N_c$ supressed because all the gluon lines are
    not in the same side of the quark line. In fact, this diagram is of order $1/N_c^2$ and is thus subdominant in the RI-loop expansion. 
    \label{Fig:quark-gluon}}
\end{figure}

Another interesting example is the quark-gluon vertex at RI-one-loop;  see Fig.~\ref{Fig:quark-gluon}. Diagram (a)
is of order $g_g$ and diagram (b) is naively suppressed by a factor $1/N_c$ respect to the tree-level contribution. In fact, the suppression is rather of order $1/N_c^2$ because the $1/N_c$ contribution involves a factor ${\rm tr}\, t^a=0$. It is, thus, subleading in the RI-loop expansion. As it was the case for the gluon self-energy, the complete set of diagrams
of order $g_g$ is obtained by
dressing the quark propagators according to
Fig.~\ref{Fig:selfquarkRainbows}. The set of diagrams of order $1/N_c^2$
is richer. Indeed, on top of dressing the quark propagators in the diagram (b) of  Fig.~\ref{Fig:quark-gluon}, we can also add infinitely
many gluon ladders between the two quark legs. It is interesting to note that these are all ultraviolet {\it finite} and, accordingly,
do not contribute to the running of the quark-gluon coupling in the ultraviolet regime.

\begin{figure}[htbp]
  \centering
  \includegraphics[width=8cm]{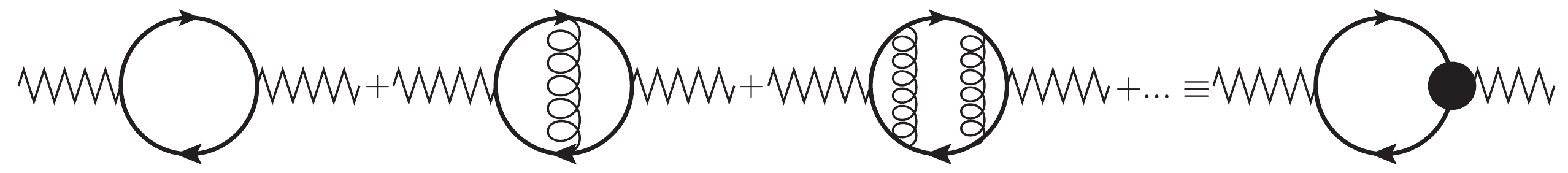}
  \caption{The infinite series of (ladder) diagrams contributing to the meson propagator at RI-one-loop order: each new rung brings additional factors $g_q^2/N_c$ from the vertices and $N_c$ from a color loop. The infinite sum can be cast in a dressed quark-antiquark-meson vertex. \label{Fig:meson}}
\end{figure}

\begin{figure}[htbp]
  \centering
  \includegraphics[width=7cm]{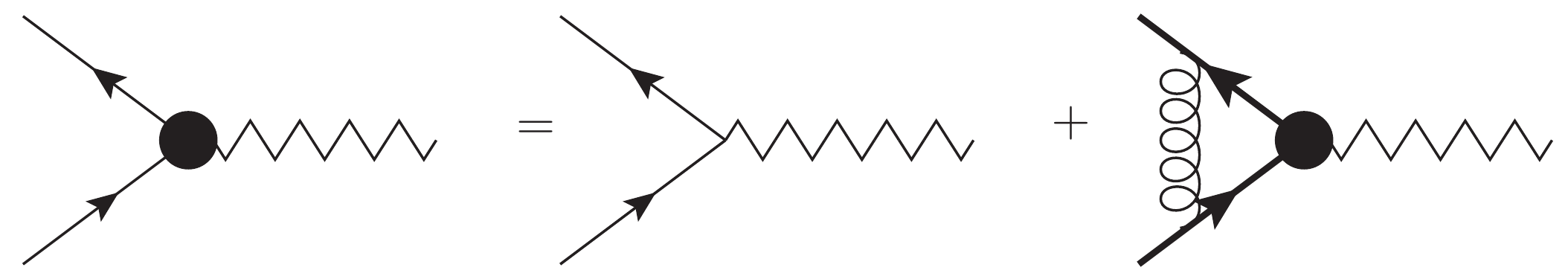}
  \caption{The linear integral equation for the quark-antiquark-meson vertex. The formally generates the infinite series of one-gluon-exchange ladder diagrams. \label{Fig:BSE}}
\end{figure}

We finally discuss the case of the meson propagator. The diagrams
contributing at RI-1-loop order are depicted in
Fig.~\ref{Fig:meson}, where it is understood that the quark
propagators are dressed according to the previous analysis; see, in
particular, Eq.~(\ref{Fig:eq_rainbow}). The meson propagator includes the infinite set of ladder diagrams and the present approximation
coincides at leading order with the rainbow-ladder approximation
for the meson spectroscopy. This infinite series can be conveniently be written in terms of a dressed quark-antiquark-meson vertex, as represented in Fig.~\ref{Fig:meson}, which satisfies the linear integral equation depicted in Fig.~\ref{Fig:BSE}. The latter clearly produces the required ladder diagrams when formally iterated in powers of the one-gluon-exchange rung. However, unlike this formal series, the integral equation for the meson vertex has a well-defined meaning and can be solved by standard (numerical) methods.

To summarize, the double expansion in $g_g$ and $1/N_c$
reproduces, at leading order, the standard rainbow-ladder
approximation. Obtaining this very powerful approximation of QCD in
the framework of a systematic expansion has three main assets.  First,
the justification of this approximation arises from a genuine analysis
of the relative values of the couplings in QCD coming from lattice
simulations.  To the best of our knowledge, the fact that the rainbow-ladder approximation can be obtained from such a systematic expansion has not
been formulated before. Second, the present analysis allows for a
precise organization of {\it subleading} corrections to the
rainbow-ladder whose contributions can, at least in principle, be
computed. This precise organization of the expansion has important
technical consequences. As we show below, it enables us to ontrol both the ultraviolet divergences and the
RG improvement of the equations (in general, a nontrivial issue for nonperturbative approximations \cite{Fischer03}) in a consistent way. Third, it
motivates the structure of the gluon propagator that has to be used in
actual calculations. In general, this requires some modeling on top
of the rainbow-ladder approximation. Here, this comes directly from
the success of the present model in the gluon/ghost sector.
We emphasize, however, that the renormalization program beyond the leading-order approximation is subtle. In fact the asymmetrical
treatment of the quark and ghost-gluon sectors may lead to the breaking of the massive version of the BRST symmetry, a symmetry that ensures the
perturbative renormalizability of the theory (\ref{eq_action}). As a consequence, the renormalization program beyond leading order may require further
work. This goes beyond the scope of the present article. 

\section{Implicit equations for the quark propagator}
\label{sec_calcul}

In this section, we  analyse in detail the quark
propagator at leading-order in the RI-loop expansion. The
integral equation depicted in Eq.~(\ref{Fig:eq_rainbow}) reads
\begin{align}
  \label{eq_rainbow2}
S^{-1}_\Lambda(p)&=-i\slashed p+M_\Lambda\nonumber\\
&+g_\Lambda^2 \int_{|q|<\Lambda} \gamma_\mu t^a
  S_\Lambda(q)\gamma_\nu t^b G_{0,\mu\nu}^{ab}(q+p),    
\end{align}
where $S_\Lambda$ represents the (unrenormalized) quark
propagator. Here, we have used an ultraviolet cutoff $\Lambda$ to regularize
possible divergences in the loop integral.

As usual, finite
correlation functions (that we note without the $\Lambda$ subscript)
are obtained by introducing renormalized fields
\beq
A_{\mu,\Lambda}^a=\sqrt{Z_A}A_{\mu}^a \quad{\rm and}\quad \psi_\Lambda=\sqrt{Z_\psi} \psi,
\eeq
and renormalized masses and coupling constant
\beq  
m_\Lambda^2=Z_{m^2} m^2,\quad M_\Lambda=Z_M M,\,\,\, {\rm and}\quad g_{q,\Lambda}=Z_{g_q} g_q.
\eeq
The renormalization factors of the quark sector can be fixed by the 
prescription
\begin{align}
  \label{eq_prescription_S}
S^{-1}(p=\mu_0,\mu_0)=-i\slashed {\mu}_0+M(\mu_0),
\end{align}
where, for short, we use the same notation $\mu_0$ for the
RG scale and for an Euclidean vector of norm $\mu_0$. We consider first a
strict version of the approximation and defer the detailed discussion
of RG effects to a subsequent section.

Equation~(\ref{eq_rainbow2}) can be decomposed in a
scalar and a vectorial component and expressed in terms of renormalized quantities. We get
\begin{widetext}
\begin{align}
\label{Aren}
  Z_\psi^{-1}(\mu_0) A(p,\mu_0)& =1
                    -Z_{g_q}^2(\mu_0)g_q^2(\mu_0)C_F\int_{|q|<\Lambda}
                    Z_\psi(\mu_0) \tilde A(q,\mu_0)\frac{f(q,p)Z_A(\mu_0)}{Z_A(\mu_0)[(p+q)^2+Z_{m^2}(\mu_0)m^2(\mu_0)]}, \\
\label{Bren}
Z_\psi^{-1}(\mu_0) B(p,\mu_0) & =   Z_M(\mu_0) M(\mu_0)+Z_{g_q}^2(\mu_0) g_q^2(\mu_0)C_F\int_{|q|<\Lambda}Z_\psi(\mu_0) \tilde B(q,\mu_0)\frac{(d-1)Z_A(\mu_0)}{Z_A(\mu_0)[(p+q)^2+Z_{m^2}(\mu_0)m^2(\mu_0)]},
\end{align}  
\end{widetext}
with
\begin{equation}
f(q,p)\equiv\frac{2p^2q^2+3(p^2+q^2)(p\cdot q)+4(p\cdot q)^2}{p^2(q+p)^2}\,.
\end{equation} 
Our notation for $A$ and $B$ (and correspondingly for $\tilde A$ and $\tilde B$) makes explicit that these functions depend on $\mu_0$ through the renormalization scale used to define the
renormalized coupling and masses. For later convenience, we have combined the renormalization factors with the associated renormalized quantities in such a way that they reconstruct the corresponding, $\mu_0$-independent, bare quantities. For instance $Z_\psi(\mu_0)\tilde A(q,\mu_0)=\tilde A_\Lambda(q)$ is independent of $\mu_0$. For $SU(N_c)$,
$C_F=(N_c^2-1)/(2 N_c)\sim N_c/2$. Accordingly, for large $N_c$,
$g_q^2C_F\sim \lambda/2$ has a finite limit.

We now discuss the renormalization of Eqs.~(\ref{Aren}) and (\ref{Bren}).  For consistency, we must treat the renormalization factors in Eqs.~(\ref{Aren}) and
(\ref{Bren}) at the order of approximation considered here, i.e., at order $g_g^0$ and
$1/N_c^0$. To this end, we
recall that the first correction to the gluon self-energy and
quark-gluon vertex are either of order $g_g$ or $1/N_c$ (see
Sec.~\ref{sec_expansion}). Consequently, $Z_A$, $Z_{m^2}$ and $\sqrt{Z_A}Z_\psi Z_{g_q}$ can all be set to 1 in Eqs.~(\ref{Aren}) and (\ref{Bren}). Next,
we observe that the integral in Eq.~(\ref{Aren}) is finite for
functions $A(p)$ and $B(p)$ behaving as the bare expressions (up
to logarithmic corrections). We can therefore consistently take
$Z_\psi$ finite. Its precise value is fixed by the condition (\ref{eq_prescription_S}) as explained below.
This generalizes the known result that, in the Landau gauge, the quark
renormalization factor is finite at one-loop order in
standard perturbation theory; see, e.g., Ref.~\cite{Gracey:2002yt}.

We are thus left with the following equations at leading order
\begin{align}
\label{eq_A}
A(p,\mu_0)&=Z_\psi(\mu_0)\nonumber\\
&-g_q^2(\mu_0)C_F\int_{|q|<\Lambda}
             \tilde A(q,\mu_0)\frac{f(q,p)}{(p+q)^2+m^2(\mu_0)}, \\
\label{eq_B}
  B(p,\mu_0) & =  Z_\psi(\mu_0) Z_M(\mu_0) M(\mu_0)\nonumber\\
  & +g_q^2(\mu_0)C_F\int_{|q|<\Lambda} \tilde B(q,\mu_0)\frac{d-1}{(p+q)^2+m^2(\mu_0)}\,.
\end{align}
The ultraviolet divergence of the momentum integral in Eq.~(\ref{eq_B}) can be 
absorbed in the bare quark mass term (first term on the right-hand side) and the renormalized equation is, consequently,
finite. It is actually more convenient to consider expressions with no
divergence at all. To do so, we compute the difference between
$B(p,\mu_0)$ and $B(\mu_0,\mu_0)=M(\mu_0)$ (note that $A(\mu_0,\mu_0)=1$), which yields
\begin{align}
& B(p,\mu_0) = M(\mu_0)+ g_q^2(\mu_0)C_F(d-1)\times\\
& \int_{q}\tilde B(q,\mu_0)\left(\frac{1}{(p+q)^2+m^2(\mu_0)}-\frac{1}{(\mu_0+q)^2+m^2(\mu_0)}\right)\,.
\label{Bsubs}
\end{align}
The integral is now finite and we can safely take the limit $\Lambda\to \infty$, if $\tilde B(q,\mu_0)$ decreases fast enough as a function of $q$ in the ultraviolet. Note also that it does not have large logarithmic
contributions as long as $p\sim \mu_0$ because the integrand is suppressed for $q\ll \mu_0$ or
$q \gg \mu_0$ as compared to the region $q\sim \mu_0$.

\section{Renormalization-Group improvement}\label{RG}

We could now try to find the self-consistent solutions of
  the previous equations which are just a particular realization of the rainbow approximation mentioned in the Introduction. However, this direct solution has
  the difficulty that the ultraviolet tails are not under control. Indeed, we observe that the integral in the right hand side
  of Eq.~(\ref{Bsubs}) involves large logarithms $\sim\ln(p/\mu_0)$ which spoil the
  validity of perturbation theory. To make this point more explicit,
  let us study the ultraviolet behavior of the solutions of Eqs. (\ref{eq_A})
  and (\ref{Bsubs}). In this regime, where asymptotic freedom holds, we
  should retrieve the results of standard perturbation theory, ie
  $A(p)\sim 1$ and $B(p)\sim (\ln p)^\alpha$ where $\alpha<0$ is given
  by an actual one-loop calculation. Instead, plugging these ultraviolet behaviors on the right-hand side of Eq.~(\ref{Bsubs}), we find that the integral behaves as $(\ln p)^{\alpha+1}$ when $p\gg m$, which is not consistent with the (assumed) behavior of the left-hand side. We
  get a clue of the origin of the problem by observing that we do
  retrieve the perturbative solution if we replace the coupling
  constant $g_q(\mu_0)$ by a running coupling constant $g_q(p)$ in
  Eq.~(\ref{Bsubs}).  The reason is now clear, for $p\gg \mu_0$, Eq.~(\ref{Bsubs})
  is not under control: even if the expansion parameters $\alpha_g$ and $1/N_c$
  are small, large logarithms spoil its validity in that regime. This is the
  standard problem of large logarithms in perturbation theory, which can be dealt with by means of the RG improvement.
  
To do so, we first make use of the RG equation:
\begin{equation}
  (\mu\partial_\mu-\gamma_\psi+\beta_{ X_i}\partial_{X_i}) S^{-1}=0
\end{equation}
where $X_i$ represents the various coupling constants and masses of the theory,
$\beta _{X_i}=\mu \partial_\mu X_i$ are the associated beta functions,
and
\begin{equation}\label{eq_gammapsi}
\gamma_\psi=\mu\partial _\mu \ln Z_\psi.
\end{equation}
This equation states
that the same correlation functions can be obtained if the
normalization prescriptions are fixed at a different scale $\mu$,
\begin{align}
  \label{eq_prescription_S2}
S^{-1}(p=\mu,\mu)=-i\slashed {\mu}+M(\mu),
\end{align}
provided that the coupling constants and masses are solutions of the
flow equations. This change of RG scale leads to a change of
normalization of the correlation function that
can be fixed by integrating the RG equation: 
\begin{equation}
S^{-1}(p,\mu,X_i(\mu))=z_\psi(\mu,\mu_0)  S^{-1}(p,\mu_0,X^0_i),
\end{equation}
with $X_i^0=X_i(\mu_0)$ and
\begin{equation}
  \label{eq:zpsi}
  \ln z_\psi(\mu,\mu_0)=\int_{\mu_0}^\mu \frac{d\mu'}{\mu'}\gamma_\psi(\mu').
\end{equation}
Evaluating now the previous equation at $\mu=p$ and using the normalization
condition Eq.~(\ref{eq_prescription_S2}), we deduce that
\begin{align}
  A(p,\mu_0)&=z_\psi^{-1}(p,\mu_0),\label{solRGA}\\
  B(p,\mu_0)&=z_\psi^{-1}(p,\mu_0)M(p). \label{solRGB}
\end{align}
We are thus left with the question of determining $z_\psi(p,\mu_0)$ and $M(p)$. To that aim, we need to change the renormalization
scale while keeping the bare quantities fixed. Of course this will simultaneously imply the running of the 
parameters in the pure gauge sector (gluon mass and couplings).
We shall first determine the functions $z_\psi(p,\mu_0)$ and $M(p)$ and then discuss the running of the remaining parameters.

\subsection{Running of $M(p)$ and expression for $z_\psi$}
 From the renormalization condition (\ref{eq_prescription_S2}) applied to Eq. (\ref{Bren}) with $\mu_0=p$, we obtain the relation
\begin{equation}
\begin{split}
& Z^{-1}_\psi(p)M(p)=Z_M(p)M(p)+Z_{g_q}^2(p)g_q^2(p)C_F(d-1)\\
& \times\int_{|q|<\Lambda} Z_\psi(p)\tilde B(q,p)\frac{1}{(q+p)^2+Z_{m^2}(p)m^2(p)}\,.
\end{split}
\end{equation}
We now take a $p$-derivative at fixed bare
quantities\footnote{ \label{footnotebare}The combinations ${Z_\psi(p)\tilde A(q,p)=A_\Lambda(q)}$, ${Z_\psi(p)\tilde B(q,p)=B_\Lambda(q)}$,
  ${Z_{m^2}(p)m^2(p)=m_\Lambda^2}$, and $Z_{g_q}(p)g_q(p)=g_{q,\Lambda}$ do not depend on $p$.} and obtain
\begin{equation}\label{eq_betaMfinal}
\begin{split}
& pM'(p)-\gamma_\psi(p) M(p)=-Z_\psi^2(p)Z^2_{g_q}(p)g^2_q(p)C_F(d-1)\\
& \times\int_{|q|<\Lambda} \tilde B(q,p)\frac{2p^2+2p.q}{[(q+p)^2+Z_{m^2}(p)m^2(p)]^2}\,.\\
\end{split}
\end{equation}
Observe that the integral in the previous equation is ultraviolet finite and we
can send the cutoff $\Lambda$ to infinity.
We finally replace $A$ and $B$ according to Eqs. (\ref{solRGA}) and
(\ref{solRGB}) and keep only the terms of order $g_g^0$ and $N_c^0$
(i.e., $Z_{m^2}=Z_\psi Z_{g_q}=1$). We then arrive at the equation
\begin{equation}\label{eq:flowM}
\begin{split}
& pM'(p)=\gamma_\psi(p) M(p)-g^2_q(p)C_F(d-1)\\
&\times \int_{q}z_\psi(q,p)\frac{M(q)}{q^2+M^2(q)}\frac{2p^2+2p.q}{\left[(q+p)^2+m^2(p)\right]^2}\,.\\
\end{split}
\end{equation}

We now derive a similar equation for $z_\psi$. The renormalization condition (\ref{eq_prescription_S2}) applied to Eq. (\ref{Aren}) with $\mu_0=p$ leads to
\begin{align}\label{eq:Z}
 Z^{-1}_\psi(p)&=1-Z^2_{g_q}(q)g_q^2(p)C_F\nonumber\\
 &\times\int_{q}
           Z_\psi(p)\tilde A(q,p)\frac{f(q,p)}{(p+q)^2+Z_{m^2}(p)m^2(p)}\,.
\end{align}
The anomalous dimension, which is needed in Eq.~(\ref{eq:flowM}), is
obtained by taking a $p$-derivative at fixed bare theory. We obtain
\begin{equation}
\label{eq_gammapsifinal}
  \begin{split}
\gamma_\psi(p)=g_q^2(p)C_F\int_{q}&\frac{z_\psi(q,p)}{q^2+M^2(q)}\Bigg[
\frac{p_\mu\frac{\partial}{\partial {p_\mu}}f(q,p)}{(p+q)^2+m^2(p)}\\ &-\frac{2(p^2+p.q) f(q,p)}{((p+q)^2+m^2(p))^2}  \Bigg]  ,
  \end{split}
\end{equation}
where, again, we have kept only terms of order $g_g^0$ and $N_c^0$ and we have used Eqs.~(\ref{solRGA}) and (\ref{solRGB}).  We note that a benefit of the present (semi)perturbative treatment is that the running coupling constant appears naturally in the flow equations (\ref{eq:flowM}) and (\ref{eq_gammapsifinal}). This plays a crucial role in obtaining the correct S\chiSB{} solutions \cite{Miransky:1984ef,Atkinson:1988mw}, which usually requires an appropriate modeling of the quark-gluon vertex in nonperturbative setups \cite{Fischer03}.

We observe that Eqs.~(\ref{eq_betaMfinal}) and (\ref{eq_gammapsifinal}) still involve $z_\psi$ and we have to relate this quatity to $M$ and $Z_\psi$
to obtain a closed system of equations. Because $Z_\psi$ is finite, we trivially obtain, from Eq.~(\ref{eq_gammapsi}),
\begin{equation}
  \label{eq_zpsirapport}
z_\psi(p,\mu_0)=Z_\psi(p)/Z_\psi(\mu_0)  ,
\end{equation}
with
\beq
\label{eq_Zpsifinal}
 Z_\psi(p)=1+g_q^2(p)C_F\int_{q}\frac{z_\psi(q,p)}{q^2+M^2(q)}\frac{f(q,p)}{(p+q)^2+m^2(p)}\,,\nonumber\\
\eeq
which is obtained from Eq.~(\ref{eq:Z}) using $Z_{g_q}(p)Z_\psi(p)=1$ and solving for $Z_\psi(p)$.
As a consequence, only functions of a single variable [$M(p)$ and $Z_\psi(p)$] have to be considered. We mention that we have a priori two different formulae
for $z_\psi$, either Eq.~(\ref{eq_zpsirapport}) or Eq.~(\ref{solRGA}). The way they are related is discussed in Appendix \ref{app:comp}.

\subsection{Running of the coupling constant and of the gluon mass}

The set of equations (\ref{eq_betaMfinal}) and
(\ref{eq_gammapsifinal}) is not closed yet because there appear the
gluon mass and the quark-gluon coupling at a running scale. In our
approximation, this can be deduced from a calculation of the
quark-gluon vertex and the gluon propagator at the same level of
approximation. This can be performed by following the procedure
described before. However for the purposes of the present paper, we will consider a simplified approximation
where the runnings of the coupling and the mass are given by simple but realistic {\it ans\"{a}tze}. We defer a more systematic analysis in the present approximation scheme to a future work.

On the one hand, the gluon mass decreases logarithmically at
large $\mu$ \cite{Tissier:2011ey}. This slow evolution is expected to have little influence
on the integrals appearing in the implicit equations
(\ref{eq_betaMfinal}) and (\ref{eq_gammapsifinal}). In the following,
we just neglect this effect and replace $m(\mu)$ by some
scale-independent value $m(\mu_0)=m_0$.

On the other hand, asymptotic freedom implies that the quark-gluon coupling $g_q(\mu)$ tends to zero in the deep ultraviolet (where all couplings have a universal running). Consequently, in this regime,
the resummed diagrams depicted in Fig \ref{Fig:quark-gluon}
simplifies greatly and we are left with the usual one-loop expression
for the beta function
\begin{equation}
\label{betafunction}
 \beta_g=-\beta_0 g^3(\mu) ,
\end{equation}
with
\begin{equation}
 \beta_0=\frac{1}{16\pi^2}\Big(\frac{11}{3}N_c-\frac{2}{3} N_f\Big) ,
\end{equation}
where $N_f$ is the number of light quarks. Equation (\ref{betafunction}) is solved as
\begin{equation}
\label{grunningsimple}
 g^2(\mu)=\frac{g^2(\mu_0)}{1+\beta_0 g^2(\mu_0)\ln(\mu^2/\mu_0^2)}.
\end{equation}
This behavior is valid as long as the RG scale is much larger than the
(quark and gluon) masses. However, there is an intermediate regime
where $\mu\gg m_0$ but where the quark-gluon coupling is still too large
to apply the usual perturbation theory. This intermediate regime could
be studied by calculating the full beta function in the RI-1-loop order, as explained above; see Fig
\ref{Fig:quark-gluon}. Instead, in this work, we use the perturbative
running and include by hand a smooth freeze-out when $\mu \simeq m_0$. Again, a more systematic treatment is deferred to a future work.
In practice, we employ the following expression for the quark-gluon
running
\begin{equation}
\label{grunning}
 g_q^2(\mu)=\frac{g^2_0}{1+\beta_0 g_0^2\ln\Big(\frac{\mu^2+x^2 m_0^2}{x^2 m_0^2}\Big)}
\end{equation}
where $x$ is a free parameter that fixes the precise
point of freeze-out. An asset of this simple truncation is that we can
vary the size of the quark-gluon vertex in the infrared and check that
S\chiSB{} occurs only for large enough coupling $g_0$.
However, we must stress
that this is an artefact of our modelization (\ref{grunning}).
We mention also that our model for the running of the coupling is such
that $g_q(\mu)$ increases with decreasing $\mu$ and saturates at $g_0$ as $\mu\to 0$. This behavior is not the one seen for instance in
fRG flows \cite{Braun:2014ata} where, the quark-gluon coupling after some dramatic increase, decreases as $\mu\to 0$.
If the decrease takes place significantly below the constituent quark mass, this effect should not have an important effect in the present analysis.
Would we treat systematically the ladder diagrams of the quark-gluon vertex,
the infrared value of the quark-gluon vertex would not be a free parameter
anymore and the variation of $g_q$ would be more realistic. This is under current investigation.
In principle, one should do the same procedure for the gluon
anomalous dimensions also, but again, we neglect this effect in the
present article.

\section{Implementation and results}
\label{sec_compare}

We now detail our numerical procedure to solve the coupled
equations (\ref{eq_betaMfinal}) and (\ref{eq_gammapsifinal}), together
with the evolution of the coupling constant (\ref{grunning}). We first
perform the angular integrals and obtain expressions where only a
one-dimensional integral needs to be performed numerically. We then
discuss the behavior of the functions $M(p)$ and $Z_\psi(p)$ when
$p\gg m$. This information is important for controlling
numerically the ultraviolet tails of the integrals. We then describe the
numerical resolution of the problem and present our results.

\subsection{Angular integration}
To simplify the study of Eqs.~(\ref{eq_betaMfinal}) and
(\ref{eq_gammapsifinal}), we first perform analytically all angular
integrals except the one over the angle $\theta$ between the vectors $p$
and $q$. Defining $u=\cos \theta$ we obtain 
\begin{widetext}
  \begin{equation}
    \begin{split}
Z_\psi(p)=  1+\frac{g_q^2(p) C_F\Omega_{d-1}}{p^2 Z_\psi(p) (2\pi)^d}\int_0^\infty dq
&\hspace{1mm} q^{d-1}\frac{Z_\psi(q)}{q^2+M^2(q)}\times\\&
\int_{-1}^1 du (1-u^2)^{\frac{d-3}{2}}\frac{2p^2q^2+3(p^2+q^2)pqu+4
  p^2u^2
  q^2}{\left(p^2+2pqu+q^2\right)\left(p^2+2pqu+q^2+ m_0^2\right)}\,,      
    \end{split}
\label{EqRendquelqA}
  \end{equation}
  \begin{equation}
    \begin{split}
      - \gamma_\psi(p)M(p)+p M'(p)  = - (d-1)\frac{g_q^2(p) C_F\Omega_{d-1}}{Z_\psi(p)(2\pi)^d}\int_0^\infty dq\
      &\hspace{1mm} q^{d-1}\frac{ Z_\psi(q)M(q)}{q^2+M^2(q)}\times
      \\
      &\int_{-1}^1 du (1-u^2)^{\frac{d-3}{2}}\frac{2p^2+2p u q}{(p^2+2q u p+q^2+ m_0^2)^2}\,,
    \end{split}
\label{EqRendquelqB}
\end{equation}
where $\Omega_{d}=2\pi^{d/2}/\Gamma(d/2)$. In integer dimensions, and
in particular in $d=4$ on which we concentrate from now on, the
integral over $u$ can be done analytically, which yields
\begin{equation}
  \begin{split}
Z_\psi(p)=  1+\frac{g_q^2(p) C_F}{32\pi^2 p^4 m_0^2 Z_\psi(p)}&\int_0^\infty dq\, q\, \frac{ Z_\psi(q)}{q^2+M^2(q)}\Big\{  \left|p^2-q^2\right|^3- m_0^4 \left[2m_0^2+3
   \left(p^2+q^2\right)\right]\\
&+\sqrt{2 q^2
   \left( m_0^2-p^2\right)+\left(m_0^2+p^2\right)^2+q^4} \left[2 m_0^4+ m_0^2
   \left(p^2+q^2\right)-\left(p^2-q^2\right)^2\right]\Big\}\,,\label{EqRenAd4}   
  \end{split}
\end{equation}
\begin{equation}
  \begin{split}
  &-\gamma_\psi(p)M(p)+p M'(p) = - \frac{3 g_q^2(p) C_F}{8\pi^2 p^2Z_\psi(p)}\int_0^\infty dq\,q \,\frac{ Z_\psi(q)M(q)}{q^2+M^2(q)}
\Bigg[ m_0^2+q^2-\frac{ m_0^4+ m_0^2 \left(p^2+2 q^2\right)-p^2
   q^2+q^4}{\sqrt{ m_0^4+2  m_0^2
   \left(p^2+q^2\right)+\left(p^2-q^2\right)^2}}\Bigg]\,.    
  \end{split}
\label{EqRenBd4}
\end{equation}
\end{widetext}
There remains to compute the angular integrals for the anomalous dimension $\gamma_\psi$ given in Eq.~(\ref{eq_gammapsifinal}).
This calculation is very similar to the one performed here for $Z_\psi$. Formally,  $\gamma_\psi$ is obtained by deriving Eq.~(\ref{EqRenAd4})
with respect to $p$ keeping the ratio $g_q^2(p) /Z_\psi(p)$ fixed on the right-hand side. 

\subsection{The ultraviolet behaviour of the equations}

Our strategy is now to look for self-consistent
solutions to Eqs.~(\ref{EqRenAd4}) and (\ref{EqRenBd4}), together
with Eq.~(\ref{grunning}). In order to do so, we shall assume specific behaviors for the functions $Z_\psi(p)$ and $M(p)$
when $p\gg m_0$ and check for their self-consistency. In the next section, we shall verify explicitly the conclusions of such an analysis by numerically solving the full system of equations.

\subsubsection{Ultraviolet limit for $Z_\psi(p)$}

We assume that $Z_\psi(p)$ behaves as some power of $\ln p$ in the ultraviolet limit ($p\gg m_0$).  We also assume that
$M(p)\ll p$ in that limit. By substituting these behaviors in Eq.~(\ref{EqRenAd4}),
it is relatively straightforward to see that the loop term is
suppressed by a positive power of $m_0^2/p^2$. Accordingly
$Z_\psi(p)\to 1$ in that limit.

\subsubsection{Ultraviolet limit for $M(p)$}

In the limit $p\gg m_0$, we find two solutions for the running mass $M(p)$. The first
one, that we call ``massive behavior'', decreases as an inverse power of $\ln p$. This is the expected
behavior away from the chiral limit. As we show
below, this solution is described by perturbation theory in the ultraviolet
limit. When the bare mass is reduced and the chiral limit is
approached, another solution appears (at least for sufficiently large
coupling constant, see below), where $M(p)$ decreases as an inverse power law in $p$. This corresponds to the
S\chiSB{} solution.

We first consider the massive case. We use that
$Z_\psi(p)\to 1$ in the ultraviolet and study the self-consistency of solutions
which behave as $M(p)\sim \ln^\alpha (p /m_0)$ at large $p$. Given that $\gamma_\psi(p)$ goes to zero as a power law in $p$, the term
including $\gamma_\psi(p)$ can always be neglected with respect to $p M'(p)$.
Consider then the integral in the right-hand side of Eq.~(\ref{EqRenBd4}) and divide it in
three parts: $q\gg p$, $m_0\ll q\lesssim p$, and $q\lesssim m_0$. The
behaviors of the integral in these three regions (together with the
logarithmic running of $g_q$) are summarized in
Table~\ref{TableUVBmassive}.
\begin{table}[ht]
    \begin{tabular}{ | c || c | c | c | p{5cm} |} 
      \hline
      $p M'(p)$ & $q\gg p$ & $m_0\ll q\lesssim p$ & $q\sim m_0$ \\ \hline
      $\ln^{\alpha-1}p$ & $p^{-2}\ln^{\alpha-1}p$ &
$\ln^{\alpha-1}p$ &$p^{-2}\ln^{-1}p$ \\ \hline
    \end{tabular}
    \caption{Large-$p$ behavior of
      Eq.~(\ref{EqRenBd4}) for the  massive solution. The first column is the assumed UV behavior of the left-hand side of the equation whereas the last three
      columns are the contributions from the different regions of integration on the right-hand side [including the prefactor $g_q^2(p)/p^2$].\label{TableUVBmassive}}
\end{table}

From this analysis, we conclude that, {\em a priori}, there are
self-consistent solutions for any value of $\alpha$ in the massive
case. We can also observe that, in the massive solution, the integral
in Eq.~(\ref{EqRenBd4}) is dominated by momenta of the order $q \sim p\gg
m_0$. This enables us to make contact with perturbation theory. Indeed,
in this regime, we can substitute the perturbative
approximation $M(q)=M_0\ln^\alpha(q/m_0)$ in the integrals.  This allows
us to approximate
\beq
 \frac{M(q)}{q^2+M^2(q)}\sim\frac{M_0\ln^\alpha(q/m_0)}{q^2}
\eeq 
and the bracket in
Eq.~(\ref{EqRenBd4}) simplifies to $2q^2\Theta(p^2-q^2)$. The integral
can now be computed easily and we get
\begin{equation}
\alpha M_0\ln^{\alpha-1}(p/m_0) = -
g^2(p)\tilde \gamma_M M_0\ln^{\alpha}(p/m_0)
\end{equation}
where $\tilde\gamma_M=3 C_F/(8\pi^2)$. By using the ultraviolet
running of the coupling constant Eq.~(\ref{grunningsimple}), we
conclude that 
\beq\label{eq:fghjk}
\alpha=-\frac{\tilde\gamma_M}{2\beta_0}.
\eeq 
One obtains the
same result as with the standard perturbative analysis. Indeed, the latter gives
\beq
\beta_M=\mu\frac{dM}{d\mu}=-M\gamma_M = -\tilde \gamma_M M g^2(\mu)+\mathcal{O}(g^4),
\eeq
whose solution is, using the perturbative running of the coupling (\ref{grunningsimple}),
\beq
\frac{M}{M_0}= \left[1+2\beta_0g_0^2\ln\left(\frac{\mu}{\mu_0}\right)\right]^{-\frac{\tilde \gamma_M}{2\beta_0}},
\eeq
in agreement with the direct analysis of Eq.~(\ref{EqRenBd4}).

Next, we want to find the ultraviolet limit of the S\chiSB{} solution. We
assume that $M(p)\sim p^\omega\ln^\delta(p/m_0)$ and repeat the
same analysis as in the massive case. We restrict the analysis to $\omega<1$ to ensure that $M(q)\ll q$ when $q\gg m$.\footnote{One can verify that in the case $\omega\geq 1$, no consistent solution can be found.} Here, one has to treat the cases $\omega$ larger, smaller or equal to $-2$ separately. In the last case, we need also to distinguish the cases $\delta$ larger, smaller or equal to $-1$. These various cases are summarized in Table~\ref{TableUVBchiSB} and we see that the only consistent solution of this type corresponds to $\{\omega=-2,\delta>-1\}$, where the dominant contribution comes from the regime\footnotemark~$m_0\ll q\lesssim p$ as expected \cite{Atkinson:1988mw,Miransky:1986ib}.

    \begin{table}[h]

    \begin{tabular}{ | c || c | c | c |}

    \hline

    $p M'(p)$& $q\gg p$ & $m_0\ll q\lesssim p$ & $q\sim m_0$\\ \hline

    $ p^{\omega>-2}\ln^{\delta}p$&

    $m^2_0p^{\omega-2}\ln^{\delta-1} p$

    &$p^{\omega}\ln^{\delta-1}p$&$p^{-2}\ln^{-1}p$ \\ \hline

    $ p^{\omega<-2}\ln^{\delta}p$&

    $m^2_0p^{\omega-2}\ln^{\delta-1} p$

    &$p^{-2}\ln^{-1}p$&$p^{-2}\ln^{-1}p$ \\ \hline

    $ p^{-2}\ln^{\delta>-1}p$&

    $m^2_0p^{-4}\ln^{\delta-1} p$

    &$p^{-2}\ln^{\delta}p$&$p^{-2}\ln^{-1}p$ \\ \hline

    $ p^{-2}\ln^{\delta<-1}p$&

    $m^2_0p^{-4}\ln^{\delta-1} p$

    &$p^{-2}\ln^{-1}p$&$p^{-2}\ln^{-1}p$ \\ \hline

    $ p^{-2}\ln^{-1}p$&

    $m^2_0p^{-4}\ln^{-2} p$

    &$p^{-2}\ln(\ln p)\ln^{-1}p$&$p^{-2}\ln^{-1}p$ \\ \hline

    \end{tabular}

    \caption{Large-$p$ behavior of Eq.~(\ref{EqRenBd4}) for the S\chiSB{} solution. The first column is the assumed UV behavior of the left-hand side of the equation whereas the last three columns are the contributions from the different regions of integration on the right-hand side [including the prefactor $g_q^2(p)/p^2$]. We have distinguished the cases $\omega>-2$, $\omega<-2$, $\{\omega=-2,\delta>-1\}$, $\{\omega=-2,\delta<-1\}$ and $\{\omega=-2,\delta=-1\}$. Only the third case yields a possible consistent solution. \label{TableUVBchiSB}}

    \end{table}

\footnotetext{ The range $0\le q\lesssim m_0$ of the integral is essentially constant and its contribution to the right-hand side of Eq.~(\ref{EqRenBd4}) is always controlled by $g_q^2(p)/p^2$. The large momentum contribution $q> p$ vanishes at $m_0=0$ [see Eq.~(\ref{eq:Litim}) below] and is thus suppressed by at least one power of $m_0^2/p^2$.}

To compute the exponent $\delta$, we can thus safely set $m_0=0$ in the integrand of Eq.~(\ref{EqRenBd4}). As before, the term in brackets becomes $2q^2\Theta(p^2-q^2)$ and, further using $Z_\psi(p)\to{\rm const}$ and neglecting $M^2(q)\ll q^2$ in the denominator of the integrand, we arrive at
\beq
\label{eq:Litim}
p M'(p) \approx - \frac{3 g_q^2(p) C_F}{8\pi^2 p^2}\int_{m_0^2}^{p^2} dq^2\,M(q)\,.
\eeq
Plugging $M(p)\propto p^{-2}\ln^\delta p$ and extracting the multiplicative constant of the leading contribution for large $p$, we find that a consistent solution requires
\beq\label{eq:fjghj}
 1+\delta=-\alpha=\frac{9C_F}{11N_c-2N_f},
\eeq
which reproduces the known results in the rainbow-ladder approximation \cite{Atkinson:1988mw,Miransky:1986ib,Fischer03,Aguilar:2010cn}. We stress that the proper implementation of the running the coupling is a key ingredient to obtain this result. The present perturbative RG treatment allows for a consistent implementation of the latter.

\subsection{Numerical implementation}

In practice, for numerical purposes, the integral over $q$ is divided in two regions, one for $q<\Lambda_1=10\, {\rm GeV}$ and the ultraviolet region for
$\Lambda_1<q<\Lambda_2=20\,{\rm GeV}$. 
In the second region the values of $Z_\psi(q)$ and $M(q)$ are replaced by their ultraviolet expressions, i.e.,
\begin{align}
\label{UVexpression1}
Z_\psi^{\text{UV}}(q)&=1\\
M^{\text{UV}}(q)&=b_0\left(\ln \frac{q^2+m_0^2}{m_0^2}\right)^\alpha +\frac{b_2}{q^2}\left(\ln \frac{q^2+m_0^2}{m_0^2}\right)^{ -(\alpha+1)} 
\label{UVexpression}
\end{align}
 where the exponent $\alpha$ is given in Eq.~(\ref{eq:fghjk}). The coefficients $b_0$ and $b_2$ are chosen in order to make $M(p)$ continuous and differentiable (so they are not free parameters).

For $p<\Lambda_1$, we sample the functions $Z_\psi(p)$ and $M(p)$ on a
regular grid with a lattice spacing of $0.05\, {\rm GeV}$. We have verified that the results presented below are converged with respect to
this choice.
We solve the self-consistent equations for the functions $Z_\psi(p)$ and $M(p)$ iteratively with initial conditions provided by their
respective perturbative expressions (\ref{UVexpression}) and (\ref{UVexpression}), with a fixed value of
$M(\Lambda_1)$.

\subsection{Chiral and massive behaviours}

\begin{figure}[h]
\centering
\includegraphics[width=\linewidth]{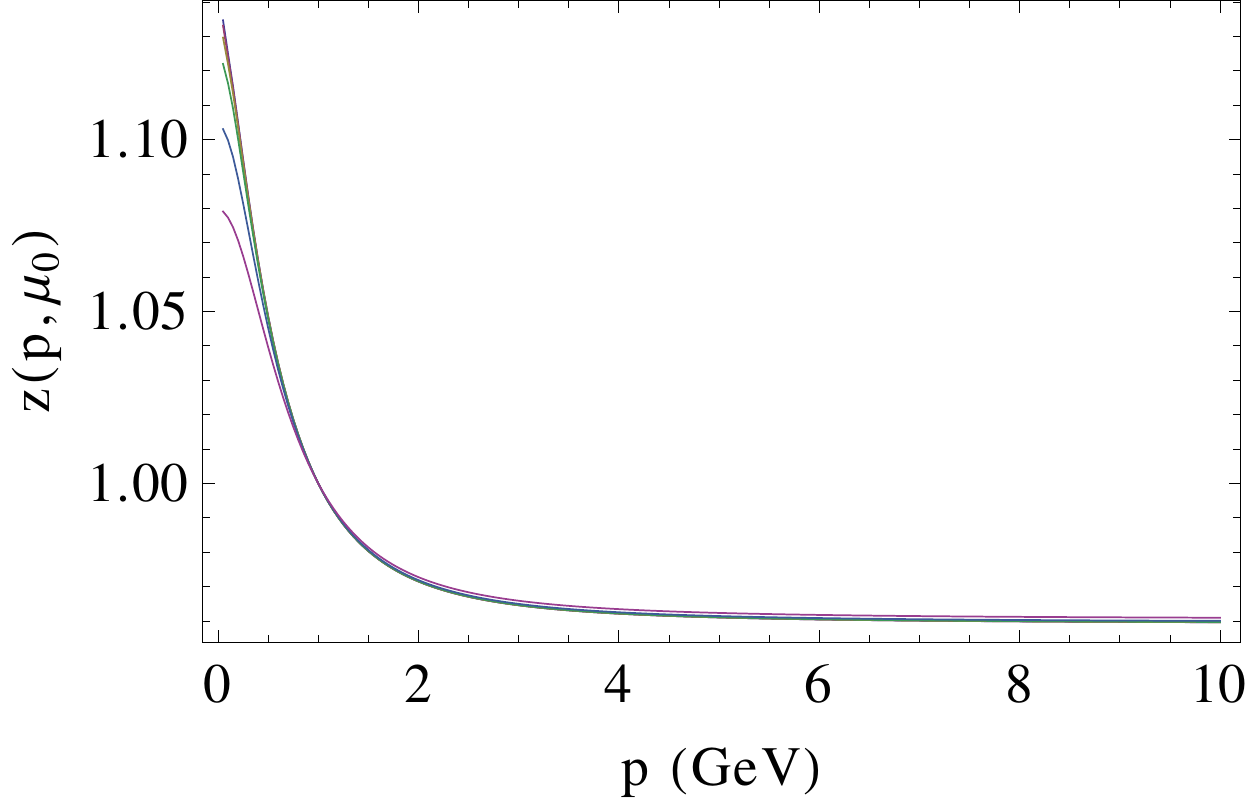}
\includegraphics[width=\linewidth]{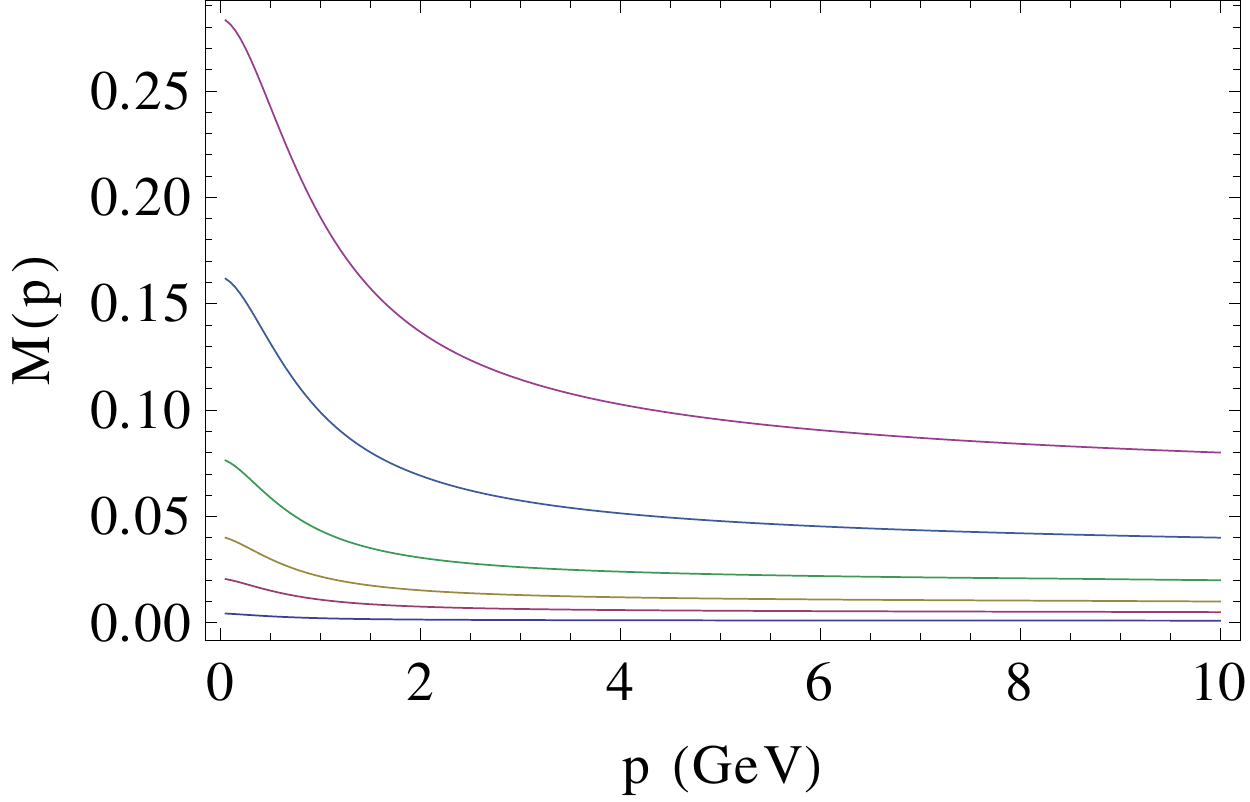}
\caption{Solutions of Eqs. (\ref{EqRenAd4}) and  (\ref{EqRenBd4}), $Z_\psi(p)$ and $M(p)$ for different values of 
$M(\Lambda_1)= 0.001, 0.005, 0.01, 0.02, 0.04, 0.08$.  
Parameters: $N_f=2$, $N_c=3$, $m_0=0.4$ GeV, $g_0=4$ and $x=5$.}
\label{fig:muestraNoChiral}
\end{figure}

\begin{figure}[h]
\centering
\includegraphics[width=\linewidth]{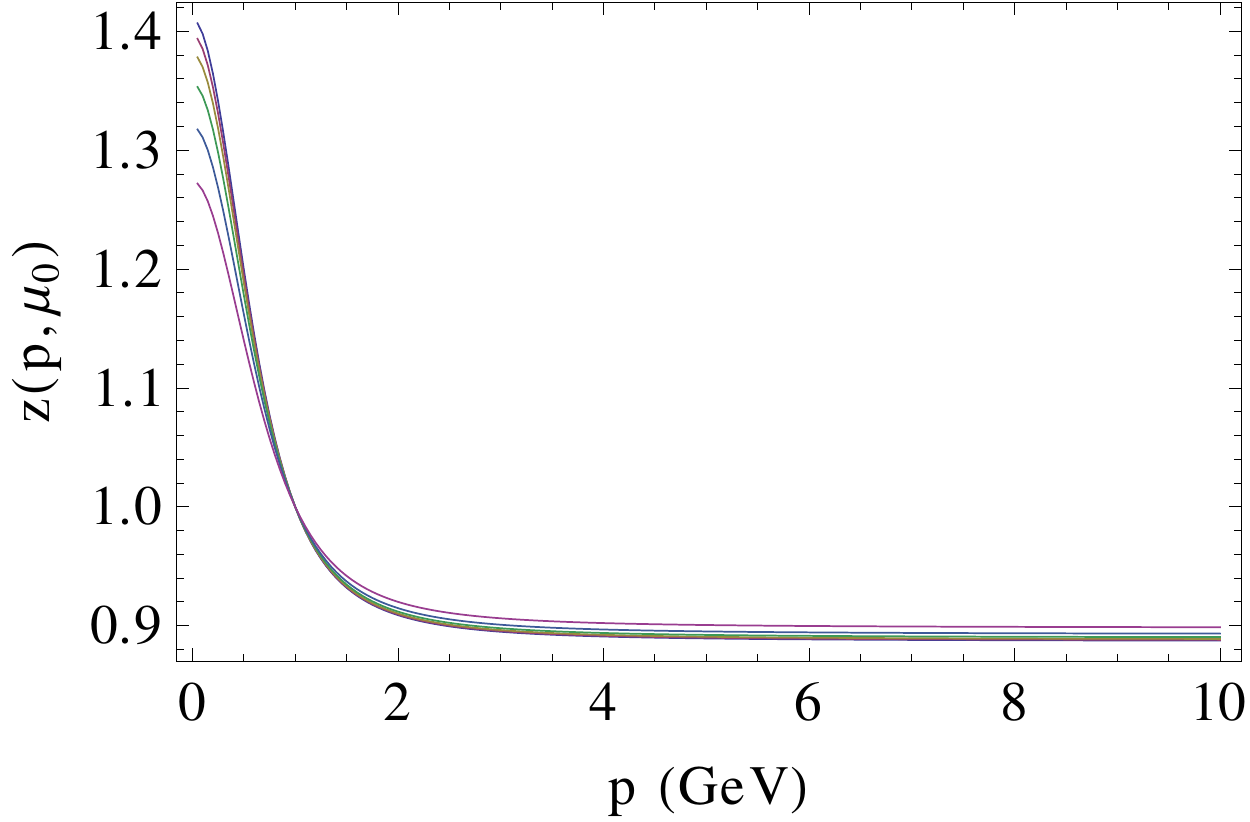}
\includegraphics[width=\linewidth]{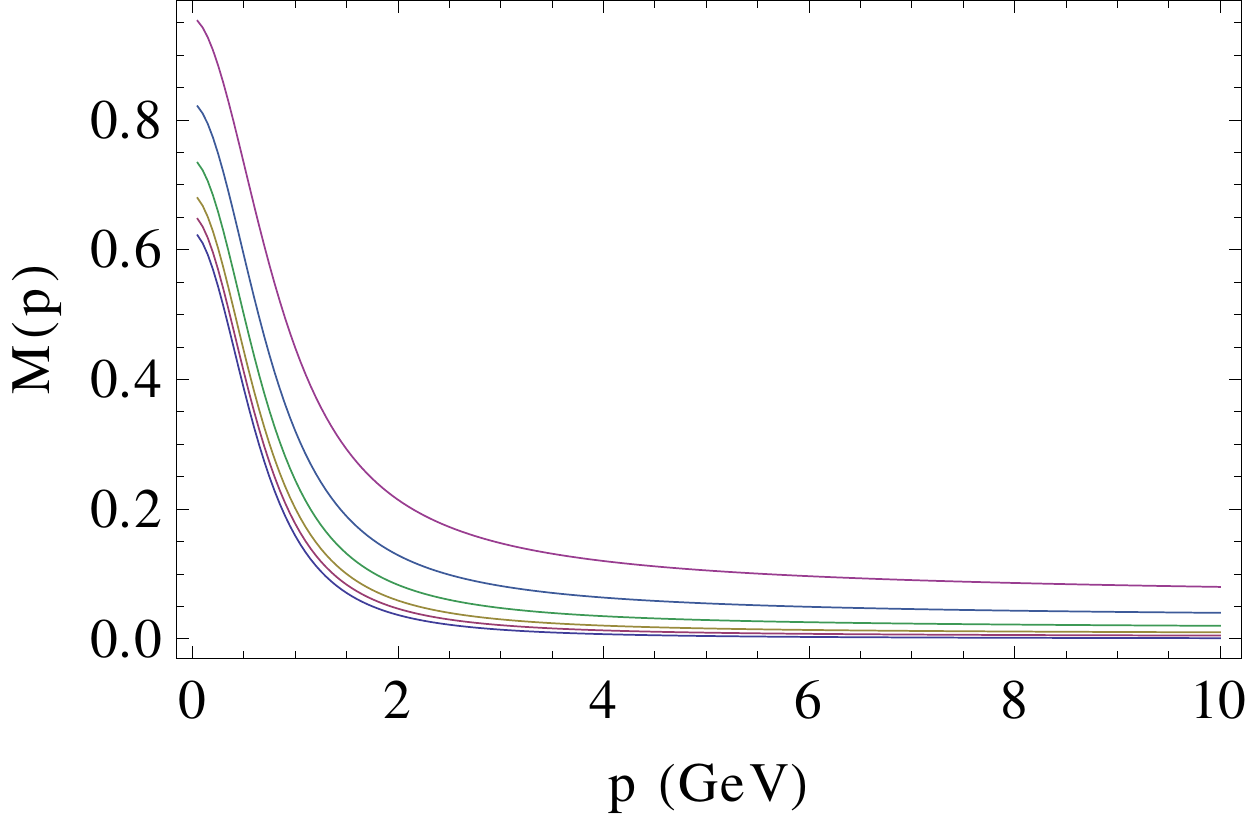}
\caption{Solutions of Eqs. (\ref{EqRenAd4}) and (\ref{EqRenBd4}),
  $Z_\psi(p)$ and $M(p)$ for different values of
  $M(\Lambda_1)= 0.001, 0.005, 0.01, 0.02, 0.04, 0.08$.  
Parameters: $N_f=2$, $N_c=3$, $m_0=0.4$ GeV, $g_0=11$ and
  $x=5$.}
\label{fig:muestraChiral}
\end{figure}

In Fig. \ref{fig:muestraNoChiral}, solutions for Eqs. (\ref{EqRenAd4})
and (\ref{EqRenBd4}) are shown for different values of
$M(\Lambda_1)$ for $g_0=4$ and $x=5$.  No chiral solution is found
for this small value of $g_0$.  However for $g_0=11$ a chiral
solution appears as shown in Fig. \ref{fig:muestraChiral}.

\begin{figure}[h]
\centering
\includegraphics[width=\linewidth]{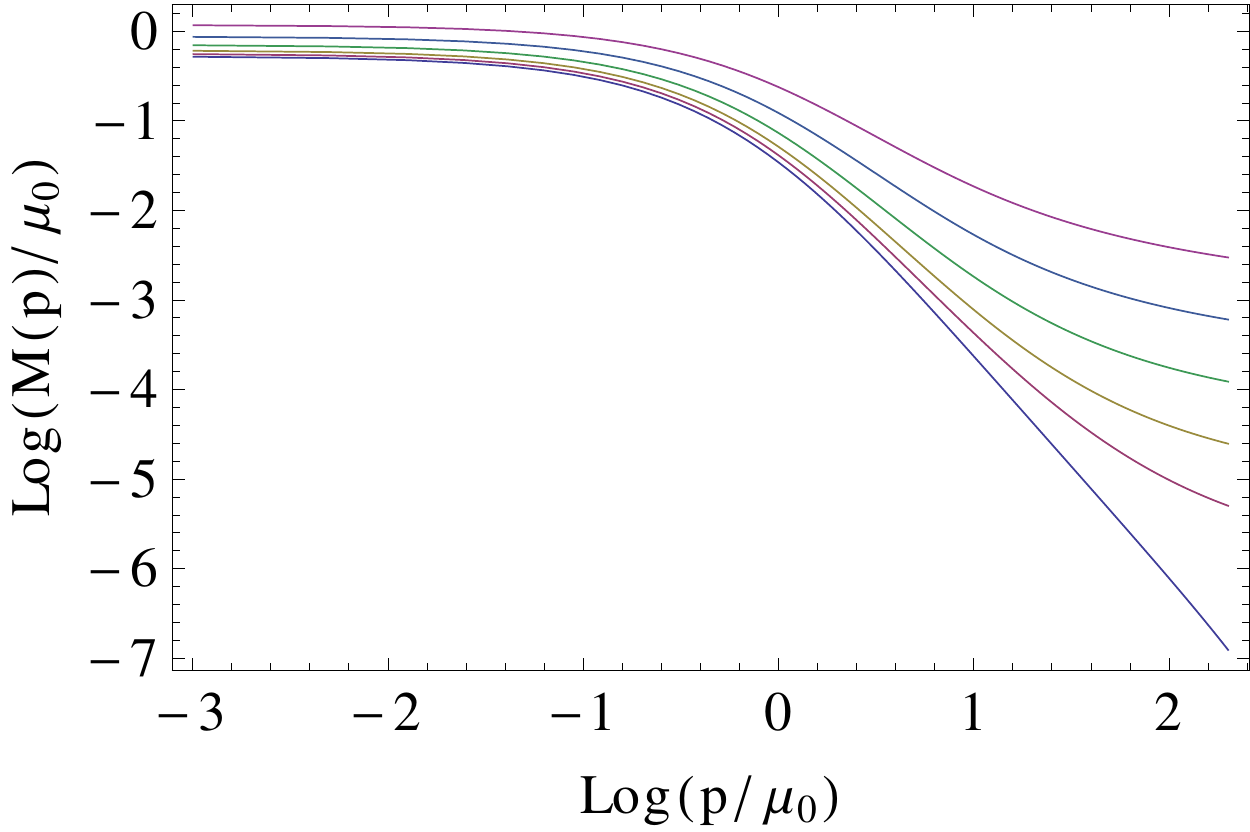}
\caption{Mass function $M(p)$ in Log-Log scale for different values of
  $M(\Lambda_1)= 0.001, 0.005, 0.01, 0.02, 0.04, 0.08$.  
Parameters: $N_f=2$, $N_c=3$, $m_0=0.4$ GeV, $g_0=11$ and
  $x=5$. We observe the onset of the power law behavior at large momentum as the chiral limit is approached. This signals the S\chiSB.\label{log-log}}
\end{figure}

\begin{figure}[h]
\centering
\includegraphics[width=\linewidth]{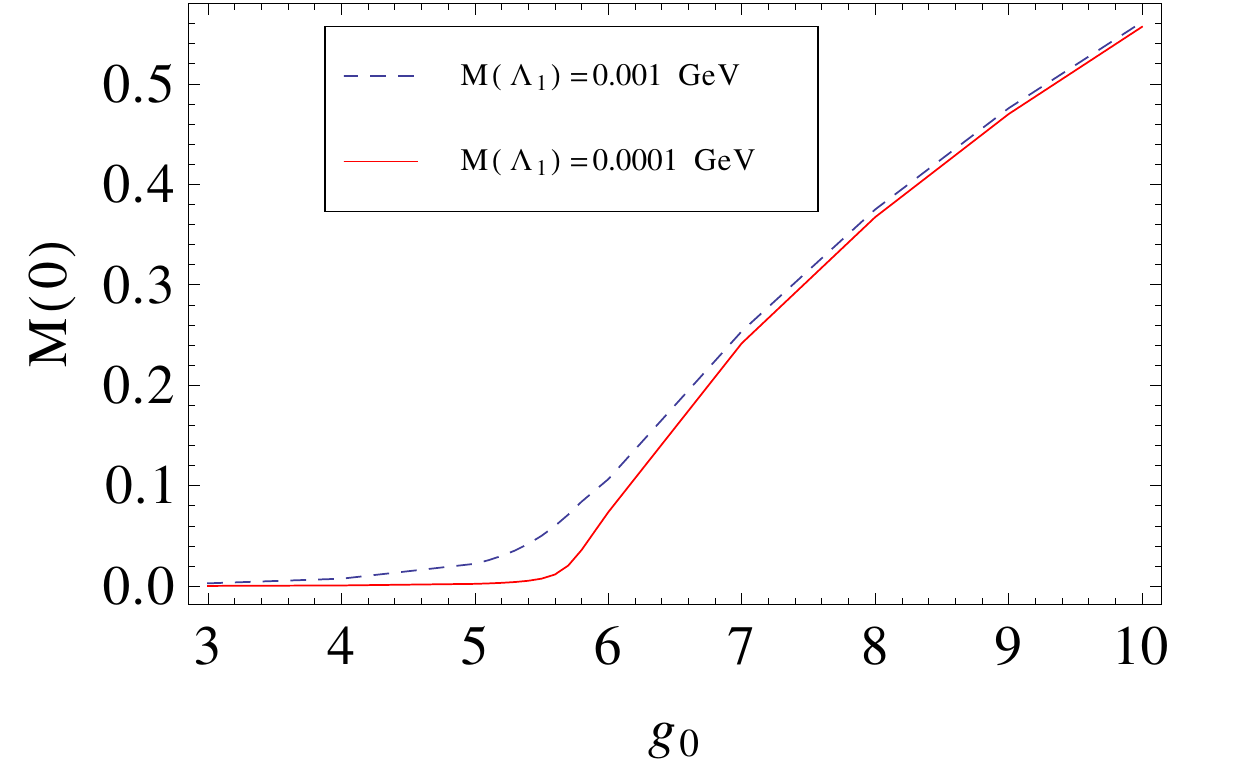}
\caption{Constituent quark mass $M(p=0)$ as a function of the coupling parameter $g_0$ for two values of the ultraviolet mass $M(\Lambda_1)$. The variation
of $g_0$ is done by keeping $\Lambda_{QCD}$ fixed.
\label{fig:regimes}}
\end{figure}

Unfortunately, in both cases the behaviour of $A(p,\mu_0)=z^{-1}_\psi(p,\mu_0)=Z_\psi(\mu_0)/Z_\psi(p)$ is not the correct one. 
This is the same problem as with the one-loop results of Ref.~\cite{Pelaez:2014mxa}. There, it was also observed
that the inclusion of two-loop corrections gave the correct shape of this function as explained in the Introduction. We expect this fuction to be better described at
RI-2-loop order. In Fig.~\ref{log-log} the mass curve $M(p)$ is represented in a Log-Log scale. One can observe the approach in the chiral limit to
an (approximate) power-law behavior.

Finally in Fig.~\ref{fig:regimes}, we illustrate the two---chirally symmetric versus chirally broken---phases of the system by plotting the
constituent quark mass $M(p=0)$ as a function of the coupling parameter $g_0$ (when varying $g_0$ we vary also $x$ in such a way to keep $\Lambda_{QCD}$ fixed). This is done for two values of the ultraviolet mass $M(\Lambda_1)$ very
close to the chiral limit. Observe that, as expected, the convergence to the chiral limit is very slow for couplings approaching the critical value.

\subsection{Comparison with lattice data for $N_f=2$}

\begin{figure}[h]
\centering
\includegraphics[width=\linewidth]{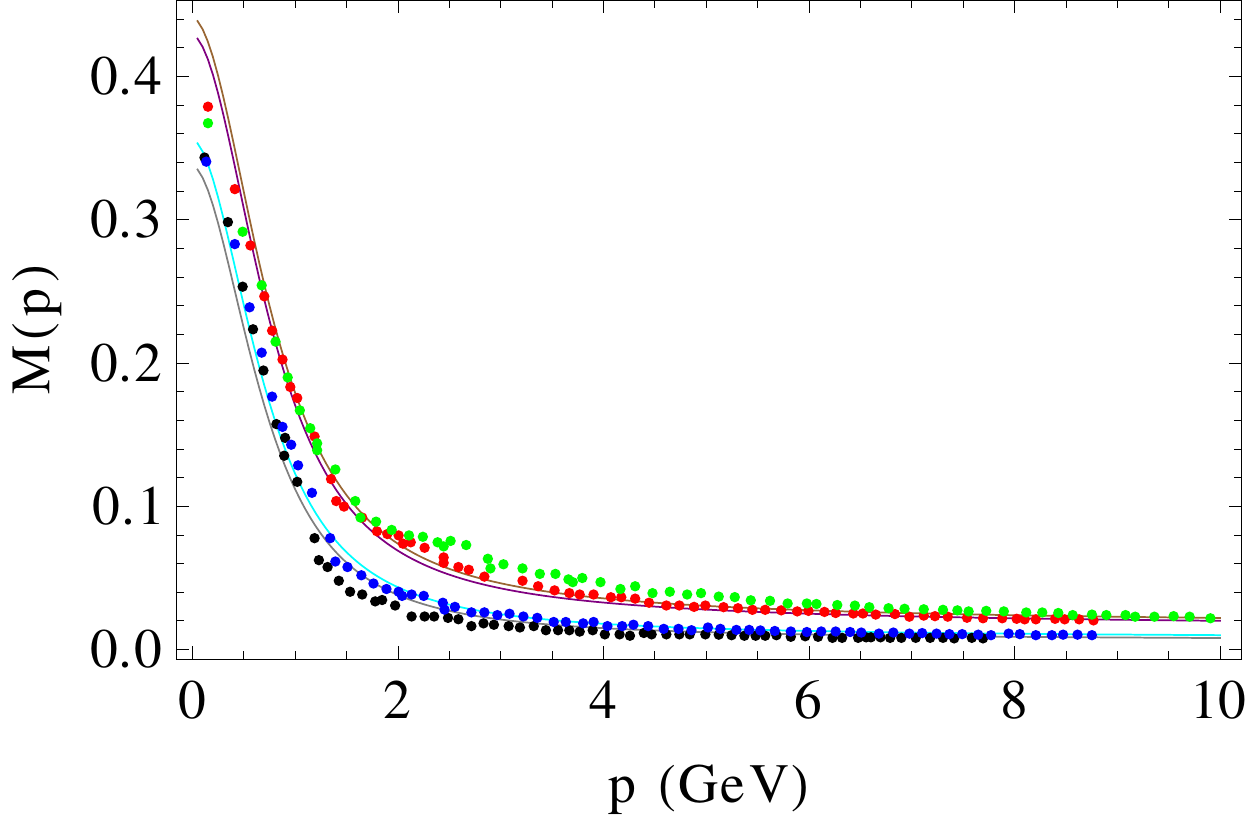}
\caption{Comparision with lattice data from \cite{Oliveira:2016muq} for $M(p)$
for $M(\Lambda_1)=0.008, 0.01, 0.02, 0.022$.
Parametres: $N_f=2$, $N_c=3$, $m_0=0.4$ GeV, $g_0=7$ and $x=5$.}
\label{fig:solmejorM1}
\end{figure}

Fig.~\ref{fig:solmejorM1} shows the comparison of the present results with lattice data from Ref.~\cite{Oliveira:2016muq}.
This is done by fitting the parameters ($g_0$, $x$) so as to
minimize the absolute error with all data-set simultaneously. After fixing those parameters,
the various curves are fitted by varying the parameter $M(\Lambda_1)$.

The agreement is quite stricking for the running mass $M(p)$. This is a qualitative improvement
with respect to the one-loop results of Ref.~\cite{Pelaez:2014mxa}. This, of course, is due to the rainbow-improvement of the one-loop expressions in the quark sector. It is, indeed,  well-known that the rainbow resummation gives good agreement with lattice data even near the chiral limit (see,
for example, Refs.~\cite{Roberts:2007jh,Aguilar:2014lha}). As explained before, the main improvement of the present work is that this resummation proceeds from a systematic expansion scheme, which allows for a consistent RG improvement of the equations.

\section{Conclusions}
\label{sec_conclusion}

We have devised a systematic expansion scheme for QCD at low energy based on a double expansion in powers of the coupling strength $g_g$ in the Yang-Mills
sector of the theory and in powers of $1/N_c$. It is based on the observation that, at low energies, the coupling $g_g$ differs significantly from the
coupling $g_q$ in the quark sector. The motivation for the $1/N_c$ expansion is more practical and allows to obtain closed expression for the various correlation functions (let us point out however that the validity of the $1/N_c$ expansion in QCD is well established in the literature, see for instance \cite{Witten:1979kh}).
At leading order, this scheme reproduces the well-known rainbow approximation. One of the benefits of our approach is however that it allows for a systematic
study of higher order corrections. Moreover, at the present leading order, we are able to implement a consistent renormalization group improvement of the
rainbow equations that yields a better control of large logarithms.

In the present work, we have considered a simplified running for the coupling. Among the possible extensions of the present work, it will be interesting to
implement a realistic renormalization group equation for the quark-gluon coupling, based on the present approximation scheme. Another interesting extension is the analysis of the next approximation order in view of improving the description of the vectorial part of the quark propagator.

The present results open the way to applications mainly in two directions. First, we would like to use the present scheme to calculate mesonic properties
such as the mass spectrum or decay rates. Given the well established success of the rainbow-ladder approximation \cite{Roberts:2007jh}, this path seems
promising. Second, we would like to explore the QCD phase diagram both at finite temperature and at finite chemical potential. The massive extension of
QCD has been already applied with success for that purpose in the heavy-quark regime \cite{Reinosa:2015oua}. The present work opens the way for the
application of this model to the lower quark masses, including the chiral limit as well as physically realistic values.

\begin{acknowledgments}

The authors would like to acknowledge the financial support from PEDECIBA program and from the ANII-FCE-1-126412 project.
NW would like to acknowledge Universit\'e Paris Diderot, where part of this work has been realized, for hospitality.
U. Reinosa acknowledges the support and hospitality of the Universidad de la Rep\'ublica de Montevideo during the late stages of this work.
Part of this work also benefited from the support of a CNRS-PICS project ``irQCD''.
\end{acknowledgments}

\pagebreak

\appendix

\section{Compatibility of the formulae for $z_\psi$}\label{app:comp}
  In the core of the text, we found two different formulas for
  $z_\psi$. In this appendix, we discuss the compatibility of these
  expressions. The first expression 
\begin{equation}\label{eq:formule}
z_\psi^{-1}(p,\mu_0)=\frac{1+g_q^2(\mu_0)C_F\int_{q}\frac{z_\psi(q,\mu_0)}{q^2+M^2(q)}\frac{f(q,\mu_0)}{(\mu_0+q)^2+m^2(\mu_0)}}{1+g_q^2(p)C_F\int_{q}\frac{z_\psi(q,p)}{q^2+M^2(q)}\frac{f(q,p)}{(p+q)^2+m^2(p)}}
\end{equation}
is obtained by replacing in Eq.~(\ref{eq_zpsirapport}) the form of
$Z_\psi$ given in Eq.~(\ref{eq_Zpsifinal}). The second expression,
obtained by combining Eqs. (\ref{solRGA}) and (\ref{eq_A}), gives
\begin{align}
z_\psi^{-1}(p,\mu_0) & =Z_\psi(\mu_0)\nonumber\\
&-g_q^2(\mu_0)C_F\int_q
            \frac{z_\psi(q,\mu_0)}{q^2+M^2(q)}\frac{f(q,p)}{(p+q)^2+m^2(\mu_0)}
\end{align}
Using the fact that, to the order at which we are computing, $Z_{m^2}=Z_\psi Z_g=1$, we can write
\begin{widetext}
\begin{eqnarray}
z_\psi^{-1}(p,\mu_0) 
            & = & Z_\psi(\mu_0)\left\{1-Z_{g_q}^2(\mu_0)g_q^2(\mu_0)C_F\int_q
            Z_\psi(\mu_0)\frac{z_\psi(q,\mu_0)}{q^2+M^2(q)}\frac{f(q,p)}{(p+q)^2+Z_{m^2}(\mu_0)m^2(\mu_0)}\right\}\nonumber\\
            & = & Z_\psi(\mu_0)\left\{1-Z_{g_q}^2(p)g_q^2(p)C_F\int_q
            Z_\psi(p)\frac{z_\psi(q,p)}{q^2+M^2(q)}\frac{f(q,p)}{(p+q)^2+m^2(\mu_0)}\right\}\nonumber\\
            & = & Z_\psi(\mu_0)\left\{1-\frac{g_q^2(p)}{Z_\psi(p)}C_F\int_q
            Z_\psi(p)\frac{z_\psi(q,p)}{q^2+M^2(q)}\frac{f(q,p)}{(p+q)^2+m^2(\mu_0)}\right\}
\end{eqnarray}

\end{widetext}
which, owing to Eq.~(\ref{eq:Z}), is nothing but Eq.~(\ref{eq:formule}). We have used that
$z_\psi(\mu_0)Z_\psi(q,\mu_0)$, $Z_{g_q}(\mu_0) g_q(\mu_0)$ and $Z_{m^2}(\mu_0) m^2(\mu_0)$ do not depend on $\mu_0$.

\end{document}